# Extreme Energy Concentration of Band-Limited Superoscillatory Vortices for Efficient Optical Micromanipulation

**Chengda Song,[1,2,3,#] Jing He,[1,2,3,4,#] Xi Xie,[5,*] Qian Wang,[4] Yijie Shen,[6,7,**] Fangwen Sun,[8,9] and Guanghui Yuan[1,2,3***]**

[1]Department of Optics and Optical Engineering, School of Physical Sciences, University of Science and Technology of China, Hefei, Anhui 230026, China

[2]State Key Laboratory of Opto-Electronic Information Acquisition and Protection Technology, Hefei, Anhui 230601, China

[3]Anhui Key Laboratory of Optoelectronic Science and Technologies, Hefei, Anhui 230026, China

[4]Institute of Materials Research and Engineering, Agency for Science, Technology and Research (A*STAR), 2 Fusionopolis Way, Innovis, Singapore 138634, Singapore

[5]School of Physics, Chengdu University of Technology, Chengdu 610059, China

[6]Centre for Disruptive Photonic Technologies, School of Physical and Mathematical Sciences, Nanyang Technological University, Singapore, Singapore

[7]School of Electrical and Electronic Engineering, Nanyang Technological University, Singapore, Singapore

[8]Chinese Academy of Sciences Center for Excellence in Quantum Information and Quantum Physics, University of Science and Technology of China, Hefei, Anhui 230026, China

[9]Hefei National Laboratory, University of Science and Technology of China, Hefei, Anhui 230088, China

[#]These authors contributed equally to this work

[*]xixie0721@163.com; [**]yijie.shen@ntu.edu.sg; [***]ghyuan@ustc.edu.cn

**Abstract**

The Abbe diffraction limit, tied to the fundamental spatial bandwidth constraint imposed by any physical aperture, remains the primary barrier to achieving ultimate far-field optical resolution and precise light–matter interactions. However, current efforts to engineer structured light fields beyond this limit often come at the cost of massive sacrifices in energy efficiency. In this work, we mathematically complete the family of non-zero azimuthal-order Circular Prolate Spheroidal Wave Functions (CPSWFs), introducing them as a complete class of band-limited superoscillatory optical vortices carrying helical phase. Compared with classical Laguerre–Gaussian (LG) beams, we rigorously prove that these eigenmodes achieve the theoretical upper bound for extreme energy concentration under strict band-limited constraints. At the scale of light–matter interactions, this optimal concentration directly amplifies the intensity gradients and angular momentum densities that govern optical forces. This advantage translates directly into a 29.9% reduction in the trapping power threshold and a 2.3-fold increase in the subdiffraction orbital rotation speed of nanoparticles. Looking forward, this fundamental physical framework not only establishes strict mathematical boundaries for structured light fields but also serves as an absolute theoretical benchmark for deep-learning inverse design, and next-generation extreme optical micro-manipulation systems.

## 1. Introduction

The Abbe diffraction limit of conventional imaging, which represents a direct consequence of the finite spatial bandwidth imposed by physical apertures, remains the primary barrier to improving far-field optical resolution and the efficiency of light–matter interactions. Efforts to engineer the point spread function (PSF) beyond this limit have followed two primary paths (alongside other strategies, such as structured illumination[1], near-field techniques[2] and nonlinear approaches[3]). The first path involves a point-by-point excitation and localization paradigm. Exemplified by confocal microscopy, stimulated emission depletion microscopy[4] (STED), photoactivated localization microscopy[5] (PALM), and minimal photon fluxes[6] (MINFLUX), this approach achieves a resolution of approximately 10 nm through fluorescent labeling. The second path leverages wavefront tailoring[7] to directly generate a subdiffraction PSF. The design methodologies for this approach have evolved from polynomial fitting and manipulating zeros[8], via Bessel-beam mode expansion[9–11], to contemporary inverse design[12] and neural networks[13,14].

However, many of these approaches either require labeling, operate outside the strict band-limit, or lack a systematic modal framework. These limitations motivate a return to the foundational constraints of the imaging system via pupil engineering, a strategy that operates strictly within band-limited constraints. The natural eigenmodes of this framework are the Circular Prolate Spheroidal Wave Functions (CPSWFs), the self-reproducing modes of finite-aperture lenses and confocal cavities, which maintain identical spatial profiles (up to scaling) before and after focusing. Pioneered by Toraldo di Francia in 1952[15] and later formalized by Slepian, Pollak, and Landau[16,17], CPSWFs are distinguished from Laguerre–Gaussian (LG), Hermite–Gaussian, and Ince–Gaussian families by their selection mechanism: while those families arise from intracavity aberrations[18–20],

CPSWFs emerge from the finite aperture size itself. Frieden[21,22], and subsequently Boivin and Boivin[23,24], showed that CPSWF-based pupil filters could generate arbitrarily small central hotspots, but computational and fabrication limitations redirected the field toward simpler pupil filters and their evaluation metrics[25,26].

The subdiffraction capabilities of these inherently band-limited modes were ultimately unlocked by a pivotal concept emerging from quantum mechanics. In 1988, Aharonov *et al*. introduced superoscillation[27], demonstrating that a band-limited function can oscillate locally much faster than its highest Fourier component. After M. Berry perceptively extended this concept into optics[28,29], Huang and Zheludev soon experimentally demonstrated such features in nanohole diffraction patterns[30] and subsequently identified the inherent superoscillatory nature of CPSWFs[31], enabling their application in far-field, subdiffraction microscopy[32,33]. Since then, the CPSWF framework[34–36] has driven successive advances ranging from label-free super-resolution confocal microscopy and 3D positional metrology of nanoparticles[37], to the recent demonstrations of limited-size object microscopy[38] and simultaneous spatial–temporal superoscillation[39].

In addition to imaging and metrology, another application of subdiffraction light fields is optical trapping and manipulation of micro- and nanoparticles, where the superoscillation could be naturally interpreted as unprecedented trapping localization and stiffness. With respect to simple scalar subdiffraction fields, structured beams can trap and rotate nanoparticles with performance far exceeding that of Gaussian beams[40-42]. Additionally, for vector fields involving engineered polarization, subdiffraction bright spots can be achieved by either azimuthally polarized (AP) or radially polarized (RP) illumination[43-46], thus enabling the formation of a stable 3D optical trap[47].

Despite these significant contributions, an interconnected modal framework is still lacking. Existing superoscillatory trapping (as well as imaging) relies on specific designs while the resulting concentration of the fields is rarely discussed, an aspect that should be of practical importance. Moreover, the theoretical framework of CPSWFs, which holds promise for providing optimal designs through its eigenvalue structures, remains incomplete in three key respects: (i) athough few have demonstrated a superoscillatory vortex through manipulation of zeros[48,49], current research has predominantly focused on zeroth-order modes with a uniform phase, leaving the natural vortex-carrying members of this family[17,50] largely unexplored; (ii) the energy concentration bound of this family has not been discussed from the aspect of optical imaging, which would provide a direct quantification of trade-off between efficiency and subdiffraction confinement; and (iii) the fundamental property of optimal energy concentration has not been systematically linked to the efficiency of light–matter interactions, nor has it been mapped to physically measurable quantities such as angular momentum densities.

In this work, we first establish CPSWFs as a complete, band-limited, and superoscillatory mode family of structured light. This family is uniquely qualified as a basis for engineering far-field super-resolution and energy concentration within a prescribed circular region. We demonstrate that $\ell$-th order CPSWF modes give rise to superoscillatory optical vortices carrying helical phase, thereby completing the family and enabling the precise construction of complex subdiffraction field structures. Furthermore, we derive a fundamental bound on the energy concentration within a finite circular region and subsequently apply this theoretical framework to the aforementioned application of optical trapping and manipulation, demonstrating that the maximal field concentration of CPSWFs translates directly into enhanced concrete performance. Specifically, this concentration enables a 29.9% reduction in the incident power threshold required for stable nanoparticle trapping and a 2.3-fold increase in the vortex-driven rotation speed along subdiffraction orbits. Ultimately, we anticipate that the CPSWF mode family will serve as a foundational tool for pushing the boundaries of optical performance in domains requiring extreme light–matter interactions, ranging from far-field microscopy to advanced optical manipulation.

## 2. Fundamentals and methods

### 2.1 CPSWF framework

CPSWFs serve as eigenfunctions of the band-limited Hankel transform, consequently, they operate as self-transforms under a rotationally symmetric 2D-Fourier transform within a finite region. This section outlines the mathematical foundations of these modes and establishes their physical relevance to the field of optics. Furthermore, it characterizes the key properties of these functions—namely, superoscillation, completeness, and optimal energy concentration—that fundamentally underpin the applications developed in Section 3.

#### 2.1.1 Definition and self-production

CPSWFs, denoted by $\psi_{n,l}^{\mathrm{FoV}}$, are defined by the eigenvalue equation of the band-limited Hankel transform[17,21,34]:

$$\int_0^{k_m} \psi_{n,l}^{\mathrm{FoV}}\left(\frac{k}{k_m}\right) J_l(kr)\, kdk = \beta_{n,l}^{\mathrm{FoV}} \psi_{n,l}^{\mathrm{FoV}}\left(\frac{r}{\mathrm{FoV}}\right);\ r \in [0, +\infty),\ \ k \in [0, k_m] \quad (1)$$

Here $r$ and $k$ are the coordinates in real space and spatial frequency space (k-space), respectively, and $J_l(\cdot)$ denotes the first-kind Bessel function of order $\ell$. The non-negative integer $n$ is the radial index, associated with the number of zeros within the band-limit—defined jointly by the maximum transverse wavenumber $k_m$ and the field of view (FoV).

To connect Eq. (1) to optical focusing, we first consider focusing with rotational symmetry. Taking $\ell = 0$, Eq. (1) reduces to the 2D Fourier transform of the field $E(k, \phi) = \psi_{n,0}^{\mathrm{FoV}}(k/k_m)$:

$$\mathcal{F}_{2D}\{E(k,\phi)\} \equiv \int_0^{k_m} \psi_{n,0}^{\mathrm{FoV}}\left(\frac{k}{k_m}\right)\left[\int_0^{2\pi} e^{ikr\cos(\phi-\varphi)} d\phi\right] kdk = \int_0^{k_m} \psi_{n,0}^{\mathrm{FoV}}\left(\frac{k}{k_m}\right) J_0(kr) kdk \quad (2)$$

where $\phi$ and $\varphi$ are the azimuthal angles in $k$-space and real-space, respectively. For $\ell = 0$, CPSWFs describe 2D distributions with band-limited spatial frequency. Our numerical procedure to construct the CPSWF follows Moumni and Karoui[51] (for more details, please refer to the Supplementary Materials SN.1). Frieden comprehensively explored the applicability of these functions as optical pupil filters[22], proving that they form an orthogonal and complete basis on both the intervals $[0,1]$ and $[0, +\infty)$ for all band-limited fields with finite energy (i.e., square-integrable fields). A detailed review on how zeroth-azimuthal-order CPSWFs achieve superoscillation has also been reported previously[35].

In particular, Eqs. (1) and (2) imply that the function's profile in $k$-space, $\psi_{n,l}^{\mathrm{FoV}}(k/k_m)$, is identical to its profile within the FoV in real-space, $\psi_{n,l}^{\mathrm{FoV}}(r/\mathrm{FoV})$, thereby establishing ***self-transforms*** under the given band limits. This property ensures that the CPSWF family is naturally selected by confocal laser cavities with finite apertures, making it as physically fundamental as the Hermite–Gaussian (HG), Laguerre–Gaussian (LG) and Ince–Gaussian (IG) families. The distinction lies in the selection mechanism: HG, LG, and IG modes emerge from intracavity aberrations (astigmatism, spherical aberration, and a combination of both), whereas CPSWFs arise from the finite aperture size itself. A discussion of how these Gaussian families converge to one another was provided in a recent work[52].

A schematic of the band-limited self-transform during full round-trip propagation, which is represented as a cascaded lens array, is shown in Fig. 1(a). Here, the pupil function with a finite radius $a$ is chosen as a properly scaled $\psi_{2,0}^{\mathrm{FoV}=1.5\lambda}$, leading to a self-transforming region of approximately $1.5\lambda f/a$ at the focal plane, shown in Fig. 1(a) and (b) respectively. Notably, non-negligible oscillation exists outside the self-transforming region, which is of great interest and is thoroughly discussed in the following paragraphs.

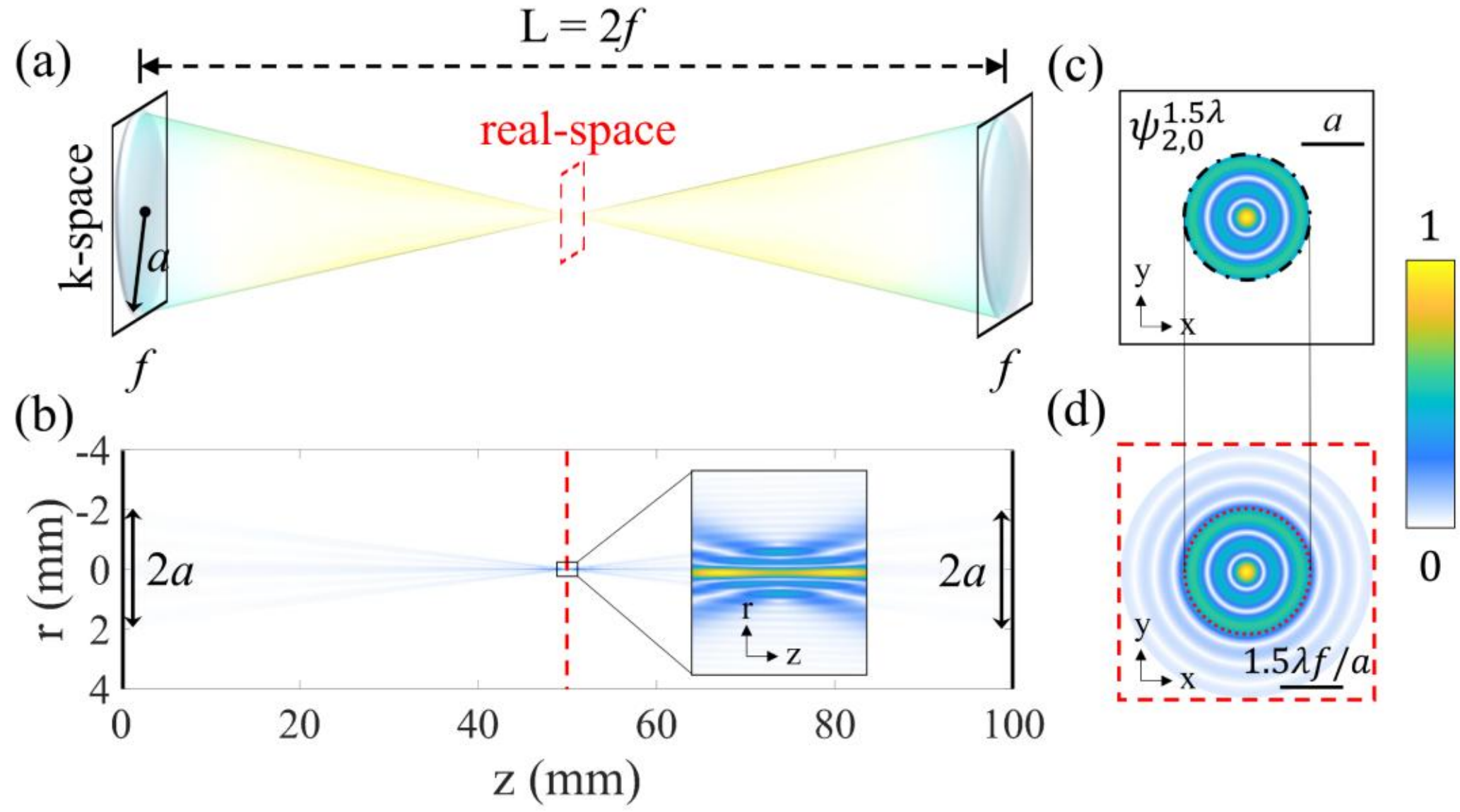

**Figure 1. Simulated full round-trip propagation of CPSWF modes in a cascaded confocal lens array. (a)** Schematic of a confocal lens pair with a finite aperture size, $a$, representing the confocal laser cavity, alongside **(b)** the amplitude profiles of the field during one round-trip propagation. **(c)(d)** Transverse amplitudes in k-space (black solid box) and real space (red dashed box). In particular, the profiles within the encircled region are identical, showcasing self-transformation behaviour. All amplitude profiles are normalized for clarity.

### 2.1.2 CPSWF optical vortices

A crucial and largely overlooked consequence of Eq. (1) is that for non-zero integer $\ell$, CPSWFs describe the band-limited optical vortices given by $E(k,\phi) = \psi_{n,l}^{\mathrm{FoV}}(k/k_m)e^{\pm il\phi}$. The 2D Fourier transform of the field yields the $J_l(\cdot)$ term in Eq. (1):

$$\int_0^{k_m} \psi_{n,l}^{\mathrm{FoV}}\left(\frac{k}{k_m}\right)\left[\int_0^{2\pi} e^{\pm il\phi} e^{i(kr\cos(\phi-\varphi))} d\phi\right] kdk = e^{\pm il\varphi} \int_0^{k_m} \psi_{n,l}^{\mathrm{FoV}}\left(\frac{k}{k_m}\right) J_l(kr)kdk \tag{3}$$

The azimuthal integral over $\phi$ is valid only for integer values of $\ell$, ruling out optical self-transforms for fields with a fractional vortex phase. Combining the radial eigenfunctions $\psi_{n,l}^{\mathrm{FoV}}(r)$ with the vortex phases $e^{\pm il\phi}$ and their associated mathematical properties (orthogonality, completeness, and square integrability), we obtain a complete, band-limited mode family. The first several modes $\psi_{n,l}^{2.0\lambda}e^{\pm il\varphi}$ (where $n = 0,1,2$, and $\ell = 0,1,2$) are shown in Figure 2(a). Band-limited, superoscillatory features deserve a prominent place in the research of optical vortices[53,54].

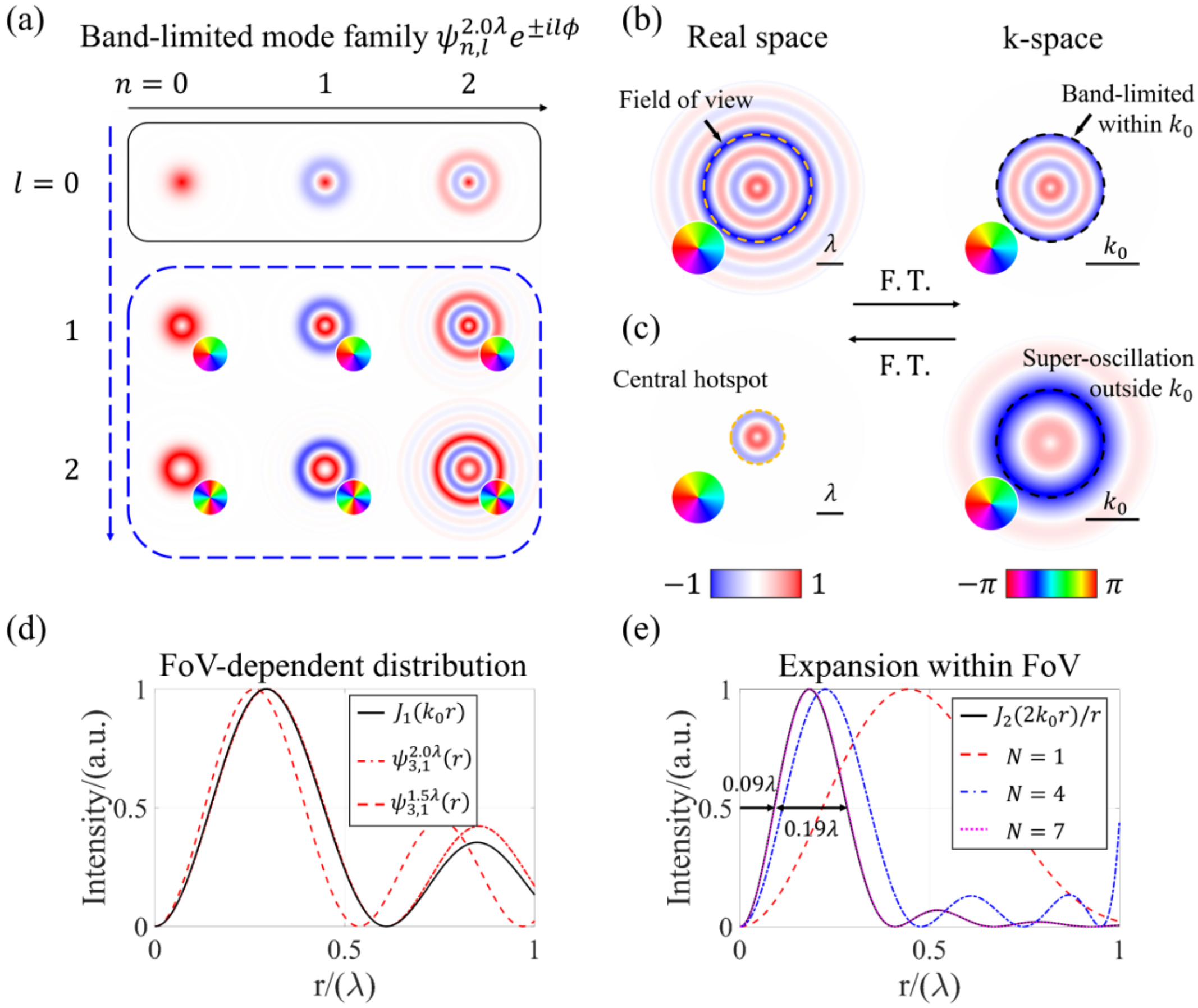

**Figure 2. Basic properties of the CPSWF optical vortex modes.** (**a**) The 2D complex amplitude profiles of the CPSWF mode family. (**b**) Superoscillatory vortex mode $\psi_{3,1}^{2.0\lambda}(x,y)e^{i\varphi}$: its full-space spectrum ($r \in [0,+\infty)$) is strictly band-limited, while (**c**) the spectrum of the central hotspot ($r \in [0,\lambda]$) exceeds the maximal transverse wavenumber $k_0$. F.T., Fourier Transform. (**d**) Radial intensities of $\psi_{3,1}^{\mathrm{FoV}}$ for varying FoVs (red solid line: FoV=2.0$\lambda$, red dashed line: FoV=1.5$\lambda$) and their highest Fourier component (black solid line). (**e**) Expansion of the target function $f_t(r) = J_2(2k_0r)e^{i\varphi}/r$ using different numbers $N$ of CPSWF modes. All curves are normalized to the same magnitude scale within the interval $[0,\lambda]$.

### 2.1.3 Band limits, halos and superoscillation

The band-limited nature of CPSWFs establishes an intrinsic link between subdiffraction hotspots and halos—intensity peaks near the FoV boundary. To elucidate this connection, we turn to $k$-space. $\psi_{3,1}^{2.0\lambda}(x,y)e^{i1\varphi}$ and its strictly band-limited Fourier spectrum are shown in Figure 2(b), where the FoV and $k_m = k_0 = 2\pi/\lambda$ are marked by yellow and black dashed circles, respectively. According to Eq. (3), the distributions within these regions are identical (up to scaling).

As first noted in Fig. 2, increasing either $n$ or $\ell$ gives rise to oscillations outside the FoV, accompanied by finer foci structures and more pronounced halos. Generating such subdiffraction features requires the Fourier spectrum to contain large amplitudes near the edge of the k-space (Fig. 2(b)), indicating that high-frequency components dominate the real-space signal. Critically, this intense high-frequency component—mandated by the self-transformation property of CPSWFs—manifests as strong, oscillatory halos surrounding the FoV. The halos are, in essence, the price the field pays to remain globally band-limited while being locally superoscillatory, establishing a fundamental trade-off between hotspot size and sidelobe strength.

While the full-space spectrum remains band-limited, Fig. 2(c) shows that truncating the central hotspot results in a spectrum exceeding $k_m$. In other words, $\psi_{3,1}^{2.0\lambda} e^{\pm il\varphi}$ "oscillates locally much faster than its highest Fourier component" [27,29], a hallmark of superoscillation that extends to optical vortex modes.

In addition, a degree of freedom absent in conventional families—the spatial bandwidth product (SBP) parameter $c \equiv \text{FoV} \cdot k_m$—allows controllable customization of superoscillation. As shown in Figure 2(d), reducing the FoV for a fixed $k_m$ reshapes the energy distributions and sharpens the subdiffraction features—a more physically transparent tuning mechanism.

**2.1.4 CPSWFs as a basis for subdiffraction field design**

The completeness of the CPSWF mode family enables the construction of arbitrary optical fields. Any smooth, square-integrable target function $f_t(r, \varphi)$ within the FoV can be expanded as follows:

$$f_{N,L}(r, \varphi) = \sum_{n=0}^{N} \sum_{l=-L}^{L} a_{n,l} \psi_{n,l}^{\text{FoV}}(r) e^{\pm il\varphi}, \quad a_{n,l} = \frac{\int_0^{\text{FoV}} \int_0^{2\pi} f_t(r,\varphi) \psi_{n,l}^{\text{FoV}}(r) e^{\mp il\varphi} r dr d\phi}{2\pi \int_0^{\text{FoV}} \left|\psi_{n,l}^{\text{FoV}}\right|^2 r dr} \tag{4}$$

The expansion coefficients $a_{n,l}$ are determined by the orthogonality of the CPSWF modes. The target function with a deep subdiffraction full width at half maximum (FWHM) of approximately $0.2\lambda$ (black curve) is shown in Figure 2(e), along with expansions for $N = 1,4,7$ (dashed curves). Any finite truncation of this expansion yields a physically realizable, band-limited field—making CPSWFs ideal for generating convergent subdiffraction target profiles in metalens optimization.

CPSWFs thus prove that hotspots can be made arbitrarily small under strict band limits—the diffraction limit is not fundamentally insurmountable. However, deep subdiffraction features carry vanishingly low energy, a consequence of the halos identified above. This intrinsic trade-off between spatial confinement and energy efficiency motivates the analysis of the optimal energy concentration below.

**2.1.5 Optimal energy concentration**

The fundamental modes (with $n = 0$) achieve the theoretical upper bound for energy concentration[55]. Specifically, for a given vortex order $\ell$, the mode $\psi_{0,l}^{\text{FoV}} e^{\pm il\varphi}$ (where $\psi_{0,l}^{\text{FoV}}$ denotes the CPSWF fundamental mode) concentrates more energy within the specified FoV than any other band-limited field of the same azimuthal order does. More precisely, a brief proof is derived from Parseval's theorem: consider a generic, smooth, square-integrable optical field with a vortex phase $e^{\pm i l_0 \varphi}$ and a band-limited spatial spectrum. By expanding this generic band-limited vortex field in the basis of the CPSWF and leveraging the self-transform property, the concentration ratio $\eta$ (defined as the energy within the FoV relative to the total energy) satisfies the following equation:

$$\eta(\mathrm{FoV}) = \frac{\int_0^{\mathrm{FoV}} \left|\Sigma \mathrm{a_n} \psi_{n,l_0}^{\mathrm{FoV}}(\frac{r}{\mathrm{FoV}})\right|^2 r dr d\phi}{\int_0^{k_m} \left|\Sigma \mathrm{a_n} \psi_{n,l_0}^{\mathrm{FoV}}\left(\frac{k}{k_m}\right)\right|^2 k dk d\phi} = \frac{\mathrm{FoV}^2}{k_m^2} \frac{\Sigma\left|a_n^2 \beta_{n,l_0}^{\mathrm{FoV}^2}\right|}{\Sigma\left|a_n^2\right|} \leq \frac{\mathrm{FoV}^2}{k_m^2} \left|\beta_{0,l_0}^{\mathrm{FoV}}\right|^2 \leq 1 \tag{5}$$

where the inequality holds since the eigenvalues $\beta_{n,l_0}^{\mathrm{FoV}}$ decay monotonically with $n$ (see Fig. 3(a) and Supplementary Materials SN.1 for details). Thus, the maximum concentration is achieved when the field consists solely of the fundamental mode $\psi_{0,l_0}^{\mathrm{FoV}}$, which possesses the largest eigenvalue $\beta_{0,l_0}^{\mathrm{FoV}}$. In addition, a generalized maximal concentration—defined by various metrics (e.g., energy density and Strehl ratio) and adaptable to different aperture shapes—has been thoroughly demonstrated in relevant numerical studies[56]. The 1D concentration bound directly accords with that theoretically given by 1D PSWFs, the efficiency limit of which is well discussed by Hou et al[57].

One might wonder whether conventional beams can achieve comparable sub-wavelength confinement. An LG mode with an expanded beam waist $w$ can indeed produce finer structures, but this comes at the cost of energy truncation at the pupil and the collapse of the self-transform property—a critical foundation for focal-field tailoring. In contrast, CPSWFs circumvent this limitation by integrating the band-limited constraint into their fundamental design from the outset.

The eigenvalues $\beta_{n,l}^{\mathrm{FoV}}$ and FWHMs for $\psi_{n,l}^{2.0\lambda}$ are displayed in Figure 3(a), which enables straightforward evaluation of the trade-off between energy concentration and subdiffraction features. The first row (corresponding to $n = 0$, the fundamental modes) represents the most concentrated modes, while higher values of $n$ yield finer features at the cost of reduced energy concentration. Modes with acceptable, practically applicable concentrations (i.e. $\eta > 0.2$) can achieve an FWHM of approximately $0.3\lambda$. Note that for larger $\ell$ or smaller FoVs, the maximum concentration decreases below. This observation is consistent with the classical result that a spatially-limited (or time-limited) and a k-limited function "cannot be very close together" [55]. Physically, this means that a band-limited optical field cannot always be perfectly concentrated within a finite spatial region.

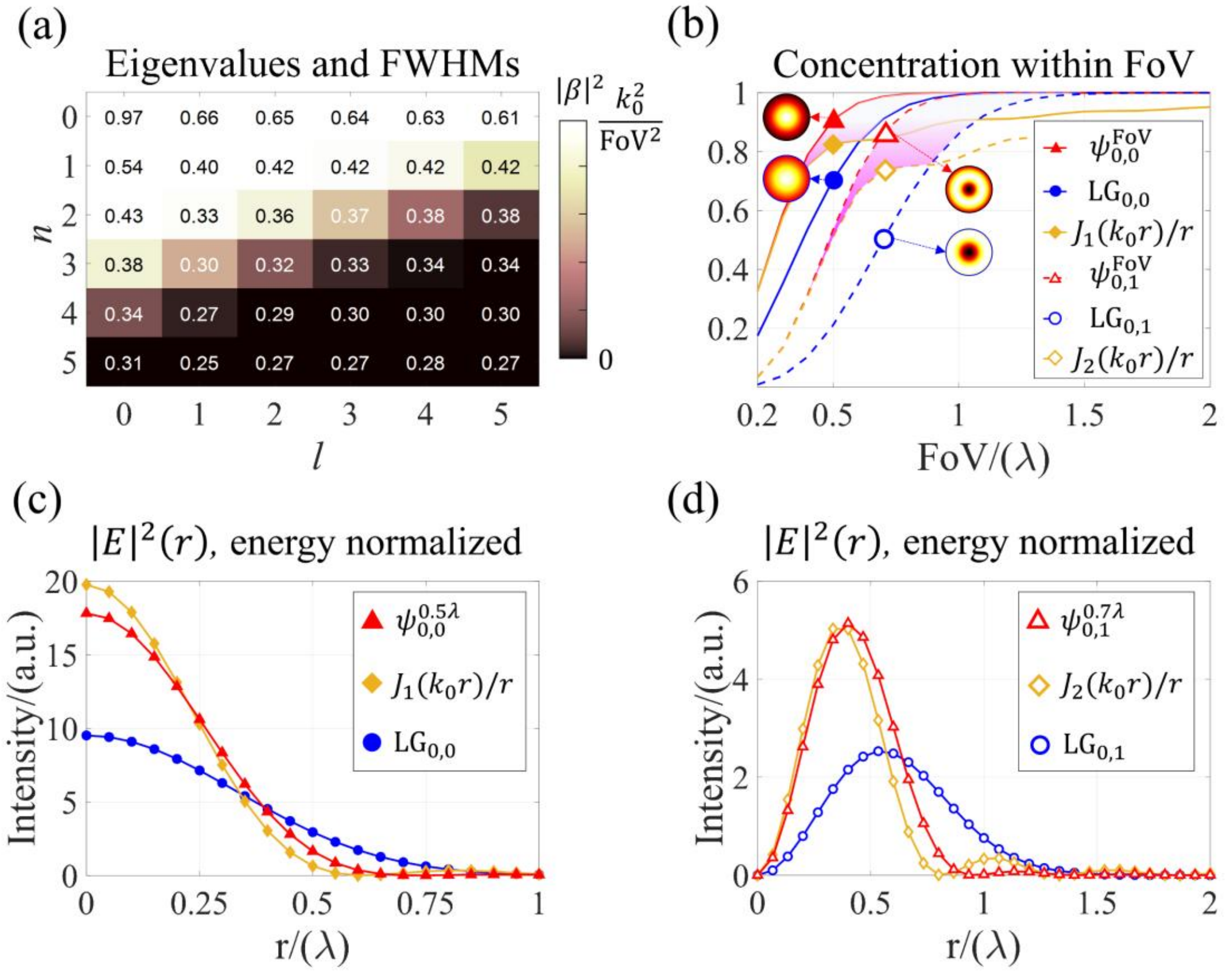

**Figure 3. Theoretical maxima of energy concentration within the FoV.** (**a**) Eigenvalues $\beta_{n,l}^{\mathrm{FoV}}$ and FWHMs for $\psi_{n,l}^{\mathrm{FoV}}$ with fixed $\mathrm{FoV} = 2.0\,\lambda, k_m = k_0$. Block color: $\left|\beta_{n,l}^{\mathrm{FoV}}\right|^2$; floating numbers: FWHM (in $\lambda$). (**b**) Energy concentration $\eta(\mathrm{FoV})$ for CPSWFs (red), LGs (blue) and Airy Disks (ADs, yellow). Solid lines: $l = 0$; dashed lines: $l = 1$. Insets: 2D field profiles of $\psi_{0,0}^{0.5\lambda}$ (red triangle), $\mathrm{LG}_{0,0}$ (blue circle), $\psi_{0,1}^{0.7\lambda}$ (red hollow triangle), and $\mathrm{LG}_{01}$ (blue hollow circle). (**c**) (**d**) Radial intensity distribution of the marked modes in panel (b). Solid markers: $l = 0$; hollow markers: $l = 1$. Here, the maximal transverse vector $k_m = k_0$ is scaled with the numerical aperture (NA) of a given imaging system.

The concentration bounds as a function of FoV and $\ell$ are compared with those of LG and Airy Disk (AD) modes in Figure 3(b). CPSWFs outperform both the LG and AD modes in the sub-wavelength regime (FoV~$0.5\lambda$, magenta region), underscoring their superiority in scenarios where a balance between energy concentration and structural refinement is required.

The radial intensity distributions of the selected modes are shown in Figures 3(c) and 3(d), respectively. For the $\ell = 0$ case (Fig. 3(c)), $\psi_{0,0}^{0.5\lambda}$ achieves $\eta_\psi = 0.923$ through sidelobe suppression at $r{\sim}0.75\lambda$, outperforming $\eta_{\mathrm{AD}} = 0.829$ and $\eta_{\mathrm{LG}_{00}} = 0.738$. This superior concentration comes with a modest trade-off: a slight degradation in both the FWHM and Strehl ratio compared with those of AD ($0.575\lambda$ vs. $0.515\lambda$, $0.90$ vs. $1$, respectively). The same sidelobe suppression mechanism is more pronounced for $\ell = 1$ (Fig. 3(d)); sidelobe suppression around $r{\sim}1.0\lambda$ yields $\eta_\psi = 0.865$, compared with $\eta_{\mathrm{AD}} = 0.739$ and $\eta_{\mathrm{LG}_{01}} = 0.501$. For $\ell = 1$ the trade-off is minimal, with only slight changes in the FWHM and peak intensity ratio compared with those of AD ($0.3947\lambda$ vs. $0.3824\lambda$, $1.02$ vs. $1$, respectively).

In short, the SBP parameter $c = \mathrm{FoV} \cdot k_m$ provides a powerful extra degree of freedom: By adjusting the FoV for a fixed $k_m$, one can optimally suppress sidelobes and achieve maximum energy concentration for any band-limited optical vortex. Furthermore, as we demonstrate in Section 3, the concentration advantage extends beyond intensity to physically measurable quantities—Poynting vector and SAM densities—and translates directly into enhanced optical forces.

## 2.2 Langevin dynamics

### 2.2.1 Time-averaged optical forces

For a spherical particle of radius $a$ satisfying the Rayleigh condition ($a \ll \lambda$), the time-averaged total lateral optical force is well predicted by the dipole approximation[58-61]:

$$\langle \mathbf{F_{opt}} \rangle = \nabla \mathrm{U} + \frac{\sigma n}{c_0} \langle \mathbf{P} \rangle - \frac{\sigma_e c_0}{n} \nabla \times \langle \mathbf{S_e} \rangle \tag{6}$$

where U is the optical potential, $\langle \mathbf{P} \rangle$ is the time-averaged Poynting vector, and $\langle \mathbf{S_e} \rangle$ is the time-averaged electric SAM density. The interaction parameters $\sigma$ and $\sigma_e$ represent light-beam cross sections, which are complex functions that are dependent on the particle properties. The constants $n$ and $c_0$ denote the refractive index of the medium and the speed of light in vacuum respectively.

Although the full expression of Eq. (6) contains eight terms—all of which are retained in our calculations—only the three shown dominate in the present configuration: the gradient force, the scattering force, and the electric curl force. The complete expression, including the chiral and magnetic contributions, is given in the Supplementary Materials SN. 3.

### 2.2.2 Langevin equation

In addition to the time-averaged optical force, the particle dynamics also include viscous drag and Brownian motion, which is governed by the Langevin equation[62]:

$$m_p \frac{d\mathbf{v}}{dt} = \langle \mathbf{F_{opt}} \rangle + \mathbf{F_D} + \mathbf{F_B} \tag{7}$$

where $m_p$ is the mass of the nanoparticle, $\mathbf{v}$ denotes its lateral velocity on the trapping plane, and $t$ is time. $\mathbf{F_D}$ and $\mathbf{F_B}$ denote the viscous dragging force and stochastic Brownian forces[63], respectively. The detailed forms of the aforementioned forces can be found in the Supplementary

Materials SN. 4.

### 2.2.3 Thermo-effects

A practical concern in metallic nanoparticle trapping is laser-induced heating[64]: absorption by the gold particle creates a local temperature gradient that drives thermophoretic drift and, at sufficiently high power, can nucleate vapor nanobubbles in the surrounding water. To avoid these uncontrolled dynamics, all simulations are conducted at incident powers of or below $25\ \mathrm{mW}$, where the thermo-effects are moderate and do not dominate the optical-force-driven dynamics.

## 3. Application in optical tweezers

Although subdiffraction trapping and rotation have been demonstrated as mentioned in the introduction, LG modes have remained the fundamental framework for tailoring forces and torques throughout the evolution of optical manipulation[65,66] mainly owing to their orthogonal vortices[67,68]—yet the mechanical properties of their band-limited counterparts, CPSWFs, and especially the relative vortices we derive, remain entirely unexplored.

The energy concentration established in Section 2 has direct physical consequences: a more concentrated field yields steeper intensity gradients and stronger local Poynting vectors, all of which govern the magnitude and spatial structure of optical forces. For CPSWFs, the additional guarantee of zero truncation loss at finite apertures means that these concentrated physical quantities are delivered to the focal plane without the efficiency penalty incurred by conventional modes. Here, we quantify these advantages by computing the time-averaged optical forces exerted by CPSWF modes on Rayleigh particles, and demonstrate two concrete results: a 29.9% reduction in the incident power required for stable trapping and a 2.3-fold increase in rotation speed along subdiffraction orbits. (Please note that, while Eq. (5) guarantees that CPSWFs outperform any band-limited alternative—including any optimized annular filters—we compare here against LG modes as the most widely used experimental benchmark.)

Throughout this section, we consider a gold (Au) nanoparticle with a radius of $50\ \mathrm{nm}$, which is carried by a water flow ($n = 1.33$) and thus is subject to a background flow velocity of $1500\ \mu\mathrm{m/s}$ at room temperature $\mathrm{T} = 293.15\ \mathrm{K}$. The incident wavelength is $800\ \mathrm{nm}$, and the focusing objective has an effective numerical aperture of $\mathrm{NA} = 1.2$. All incident beams are left-circularly polarized (LCP), which preserves rotational symmetry and only slightly affects the modal properties (e.g. FWHM, energy concentration) under high-NA focusing. In this regard, a systematic treatment of vectorial self-transforms and the degradation of modal properties under high NA is provided in the Supplementary Materials SN. 2.

To replicate realistic trapping conditions as closely as possible, we perform finite element method simulations (COMSOL Multiphysics) to verify the claimed trapping advantages. We first compare the trapping performance of the zeroth-order CPSWF mode $\psi_{0,0}$ against that of the conventional $\mathrm{LG}_{0,0}$ mode. Stable optical trapping requires two conditions. From the force perspective, the maximal optical restoring force must overcome the drag exerted by the carrying flow. From the energy perspective, the effective potential $\mathrm{U_{eff}}$ must form a local minimum, deep enough to confine the particle against Brownian motion, whose energy scale is approximately $10\ \mathrm{k_B T}$ (where $\mathrm{k_B}$ is the Boltzmann constant)[63,69].

The optical potential landscapes and force fields generated by $\mathrm{LG}_{0,0}$ and $\psi_{0,0}$ at an incident power of $11\ \mathrm{mW}$ (normalized to the pupil area $A_p$, $0.367\ \mathrm{mW/mm^2}$ in power density considering a typical $A_p \sim 30\ \mathrm{mm^2}$) are compared in Figures 4(a) and 4(b). The potential well produced by $\psi_{0,0}$ is clearly more localized than that produced by $\mathrm{LG}_{0,0}$, with a sharper peak and narrower FWHM ( $-115.89\ \mathrm{k_B T}$ vs. $-90.55\ \mathrm{k_B T}$ and $408.73\ \mathrm{nm}$ vs. $452.90\ \mathrm{nm}$ , respectively). Forces induced by both modes mainly consist of a radial component since the spin–orbital interactions are negligible; thus, the gradient forces—directly related to field intensities—dominate. The forces along the y-direction together with the effective potential (inset) are shown in Fig. 4(c), and the corresponding trapping stiffnesses are $\kappa_\psi = 1.349\ \mathrm{pN \cdot (\mu m \cdot mW)^{-1}}$ and $\kappa_{\mathrm{LG}} = 0.867\ \mathrm{pN \cdot (\mu m \cdot mW)^{-1}}$. At this power level, $\psi_{0,0}$ satisfies both trapping criteria, whereas $\mathrm{LG}_{0,0}$ does not. Conversely, the same trapping performance can be achieved by using $\psi_{0,0}$ with a 29.9% reduction in incident power.

This theoretical assertion can be verified by Langevin dynamics simulations. Representative nanoparticle trajectories under the two fields are shown in Figures 4(d) and 4(e), respectively. Under $\mathrm{LG}_{0,0}$ illumination, particles experience noticeable retardation in the encircled region but ultimately escape because of the combined effects of dragging and Brownian fluctuations. In contrast, under $\psi_{0,0}$ illumination, the particles remain stably confined within the potential minima. The resulting spatial distribution of trapped particles is shown in Figure 4(f); the inset shows the probability density function (PDF) along the x- and y-directions.

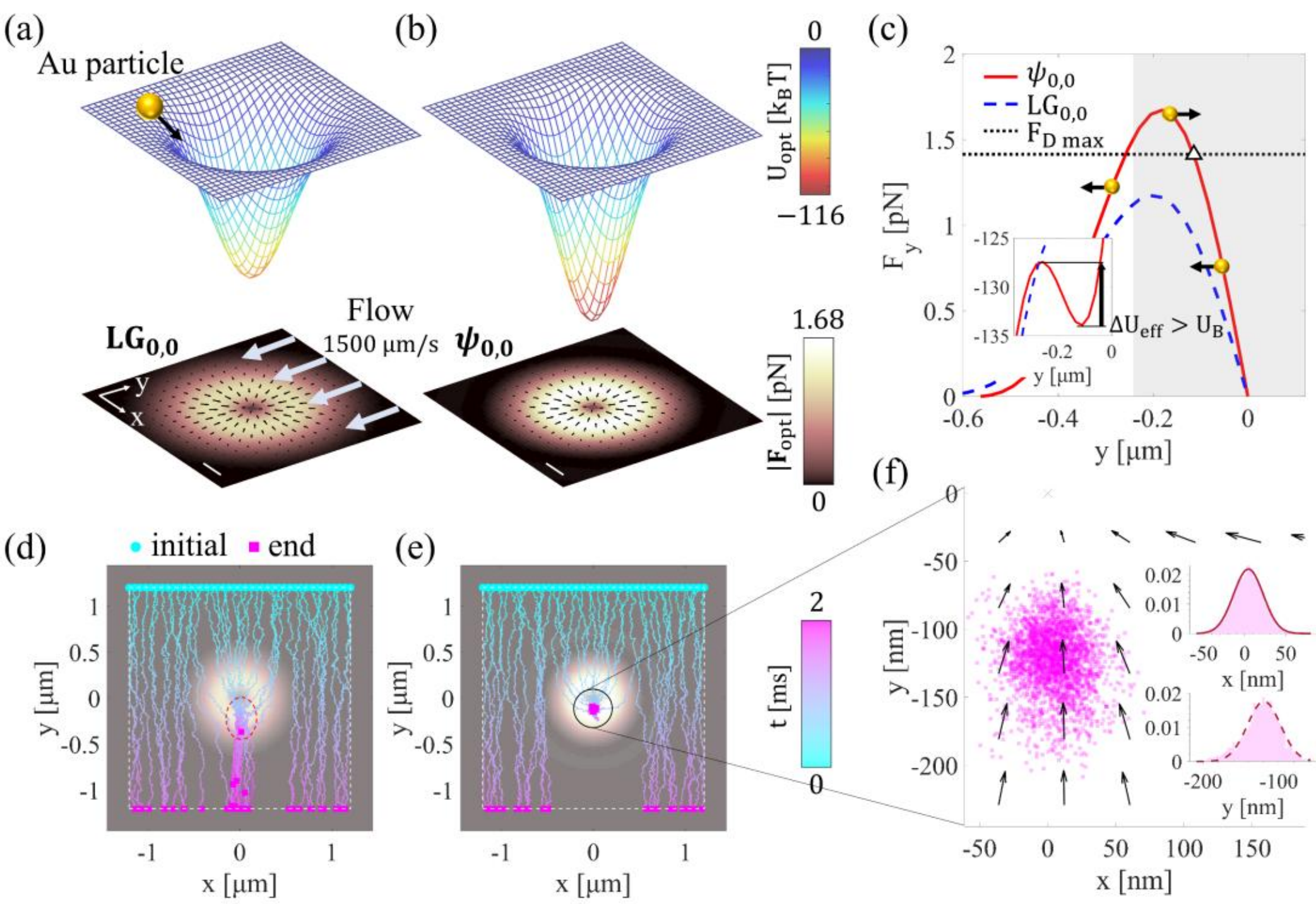

**Figure 4. Comparison of optical potential, total force distributions, and nanoparticle trapping performance between the $\mathbf{LG_{0,0}}$ mode and the $\boldsymbol{\psi_{0,0}}$ mode.** (**a**) (**b**) Optical potential and force distributions generated by the incident $LG_{0,0}$ and $\psi_{0,0}$ modes, respectively. Au nanoparticles with a radius of $50\ \mathrm{nm}$ are subject to a background flow velocity of $1500\ \mu\mathrm{m/s}$. The incident wavelength is $800\ \mathrm{nm}$. Scale bar: $150\ \mathrm{nm}$. (**c**) Force along the y-direction; $\psi_{0,0}$ incidence overcomes the maximal dragging force by approximately $1.42\ \mathrm{pN}$. The particles within the shaded area are driven toward the equilibrium points (black triangles). On the other hand, the local potential well should be deep enough to hold within Brownian motion; inset: effective potential $U_{eff}$. (**d**) (**e**) Simulated nanoparticle trajectories under $LG_{0,0}$ and $\psi_{0,0}$ illumination, respectively. The particles exhibit obvious retardation in the red-dashed circle but are stably trapped within the black-solid circle. Dashed squares in (**d**) (**e**): simulation boundaries. The cyan dots and magenta squares represent the initial and end positions of the particles, respectively. (**f**) Zoomed-in view of the spatial distribution of trapped particles and total optical forces; inset: Gaussian-fitted PDFs along the x- and y-directions, where $(\mu_x, \sigma_x) = (-4.4\ \mathrm{nm}, 18.1\ \mathrm{nm})$ and $(\mu_y, \sigma_y) = (-119.4\ \mathrm{nm}, 22.0\ \mathrm{nm})$. The distribution is distorted along the -y direction, according to the dragging force.

One might argue that widening the waist of the LG beam on the pupil plane could similarly narrow the trapping potential. However, as discussed in Section 2, this inevitably introduces non-negligible energy truncation at the pupil edge—a significant energy loss that cannot be ignored when energy efficiency is the figure of merit. In contrast, the intrinsic band-limited nature of CPSWFs guarantees zero energy truncation at the aperture, embodying the simultaneous concentration in both k-space and real-space.

We next examine the rotational dynamics induced by the first-order vortex mode $\psi_{0,1}e^{+i\phi}$.

Unlike in the trapping scenario, where the background flow velocity constitutes the dominant competing force for examining the trapping ability, we focus exclusively on the azimuthal orbiting and therefore omit the background flow. The optical potential and force field at an incident power of $25\ \mathrm{mW}$ are shown in Figure 5(a). At radial equilibrium (marked by the white dashed line in Fig. 5(a)), the azimuthal force—directly responsible for driving rotation—reaches approximately its maximum and thus dominates. The ratio of the azimuthal force to the radial force can be further enhanced by increasing either the particle radius or the topological charge $\ell$. In this work, we choose a relatively low force ratio to stabilize the radial dynamics.

The rotation of the nanoparticles under $\psi_{0,1}$ and $\mathrm{LG}_{0,1}$ illumination at equal incident powers (phase terms omitted hereafter) are shown in Figure 5(b). The corresponding phase–space diagrams, referenced to the origin, are shown in Fig. 5(c). Trajectories involving 41 particles in total are collected at the same initial position and spanning of a duration of $5\ \mu\mathrm{s}$. The radial PDF (inset) reveals that $\psi_{0,1}$ confines the particle to a mean orbital radius of $281.6\ \mathrm{nm}$ with a standard deviation of $14.8\ \mathrm{nm}$, both smaller than those observed under $\mathrm{LG}_{0,1}$ $(328.8 \pm 21.7\ \mathrm{nm})$. Critically, both the average and standard deviation of the $\mathrm{LG}_{0,1}$ orbital radius are greater than those of $\psi_{0,1}$, indicating finer structural confinement and greater radial trapping stiffness under the same power.

To elucidate the physical origin of the enhanced rotation, Fig. 5(d) displays the time-averaged angular velocity $\omega$ as a function of incident power. At a selected incident power of $25\ \mathrm{mW}$, $\psi_{0,1}$ yields an approximately 2.3-fold increase in angular velocity compared with $\mathrm{LG}_{0,1}$ $(1.96\ \mathrm{rad/ms}$ versus $0.87\ \mathrm{rad/ms})$. This enhancement can be understood from the angular momentum landscape shown in Fig. 5(e): although both beams carry the same total OAM of $1\hbar$ per photon, $\psi_{0,1}$ concentrates a higher longitudinal OAM density $L_z$ at the trapping ring radius, resulting in a stronger local azimuthal momentum transfer to the particle. The SAM density $S_z$ is also slightly redistributed but remains secondary to the orbiting of the driving particles around the beam axis. Together, these angular momentum distributions provide a concise physical picture—the faster rotation under $\psi_{0,1}$ illumination is a direct consequence of the laterally concentrated orbital momentum density enabled by the CPSWF family.

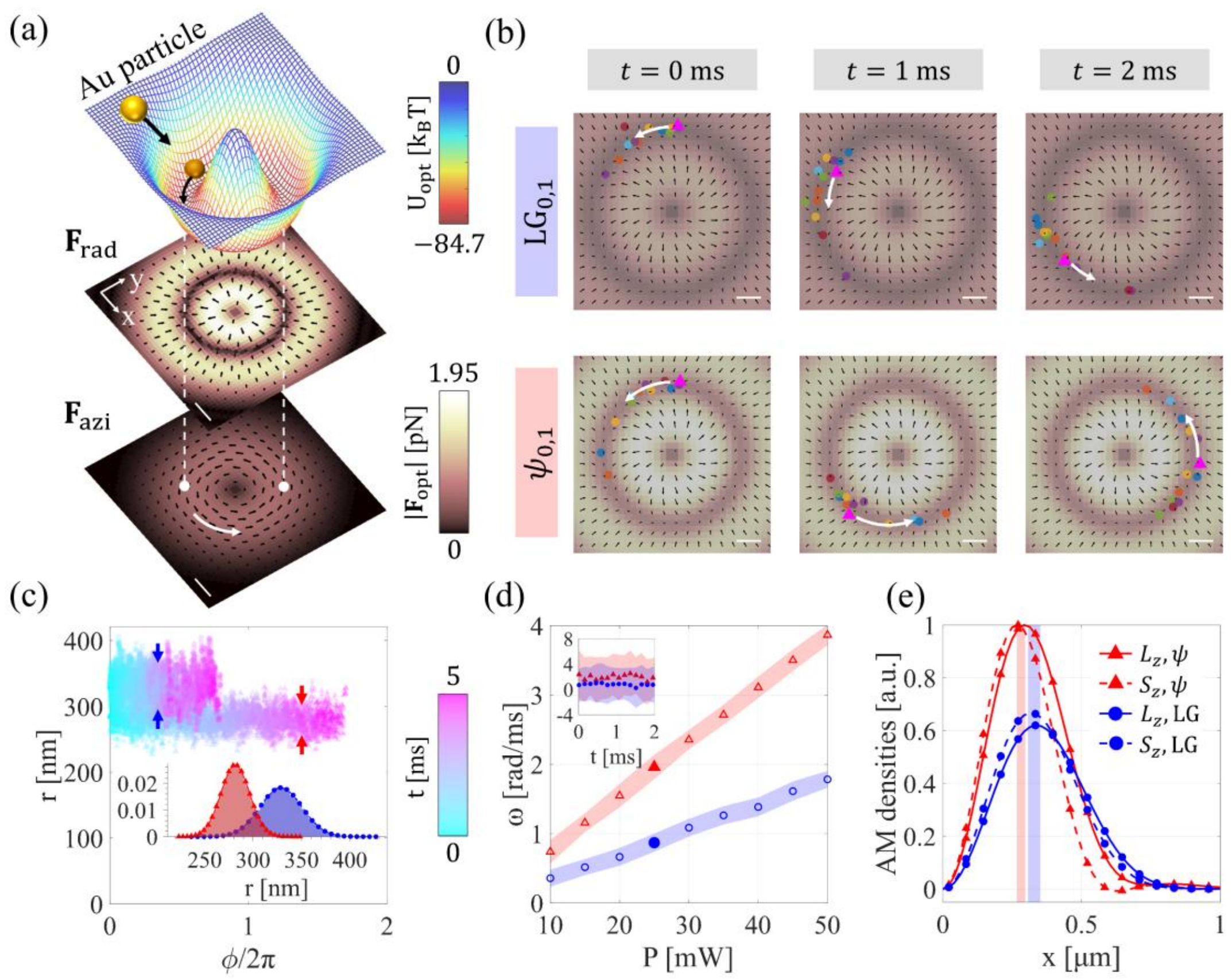

**Figure 5. Comparison of nanoparticle rotation performance between the $\mathbf{LG_{0,1}}$ and $\boldsymbol{\psi_{0,1}}$ modes.** (**a**) Optical potential and force field of the $\psi_{0,1}e^{+1i\phi}$ mode. White dashed line: radial equilibrium. The radial force is approximately 2 times larger than the azimuthal force, which is sufficient to freeze the radial dynamics. Scale bar: 150 nm. (**b**) Simulated rotation of trapped particles under $\mathrm{LG}_{0,1}e^{+1i\phi}$ (top) and $\psi_{0,1}e^{+1i\phi}$ (bottom) illumination. The particle marked as magenta triangles allows us to directly view the degree of rotation. Timestamps are relative to the time when data collection began. Scale bar: 100 nm. (**c**) Phase-space diagrams referenced to the orbital centers; inset: radial PDF ($\psi_{0,1}$ in red, $\mathrm{LG}_{0,1}$ in blue). (**d**) Time-averaged angular velocity ($\psi_{0,1}$ in red triangles, $\mathrm{LG}_{0,1}$ in blue circles) as a function of incident power. Filled markers: selected incident power of 25 mW, where laser-induced heating is negligible. Inset: instantaneous angular velocity at 25 mW, showing the fluctuation under Brownian motion and Stokes viscous drag. Shaded regions represent one standard deviation. (**e**) Line scans along the x-axis of the normalized OAM densities $L_z$ (solid) and SAM densities $S_z$ (dashed) for $\psi_{0,1}$ (red) and $\mathrm{LG}_{0,1}$ (blue), respectively. The vertical shaded bands mark the trapping ring.

From a practical standpoint, the required pupil functions are readily realizable using custom-designed q-plates, spatial light modulators (SLMs) or metalenses, which provide full dynamic tunability but at the cost of intrinsic modulation losses. The core advantage of the numerical results presented above is that they establish theoretical performance bounds for optical trapping under strict band-limited constraints, serving as an absolute theoretical benchmark against which future implementations can be evaluated.

## 4. Conclusion

This work establishes Circular Prolate Spheroidal Wave Functions as a complete, band-limited mode family for structured light, uniquely suited to finite-aperture systems. By recognizing that non-zero azimuthal orders yield superoscillatory optical vortices and that the fundamental mode of each order achieves the provable maximum energy concentration (Eq. 5), we redefine the diffraction limit not as a barrier but as a trade-off parameterized by the spatial bandwidth product $c = \mathrm{FoV} \cdot k_m$—one that can be quantified, tuned, and ultimately optimized for practical applications. This concentration advantage propagates to intensity gradients, Poynting vectors, and angular momentum densities, and manifests directly in optical forces. In the optical tweezer configuration investigated herein, the superior energy concentration of the CPSWFs enables a 29.9% reduction in the incident power required for stable nanoparticle trapping, along with a 2.3-fold increase in the rotation speed along a narrower orbit. The simulation results presented herein establish achievable performance benchmarks for band-limited finite-aperture trapping systems, providing a clear theoretical foundation for practical implementation.

We contend that CPSWFs occupy a unique niche in the field of structured light: they are the intrinsic eigenmodes naturally selected by finite-aperture systems. It is anticipated that this mode family will find widespread applications across a broad range of research areas that have long sought to push the boundaries of optical performance, from far-field microscopy and metalens design to the engineering of optical forces, angular momentum, and other light–matter interaction effects. More generally, wherever the physics imposes a spatial bound—such as a finite pupil, bandwidth limitation, or spatial confinement—are imposed, the key question is not whether the diffraction limit can be "broken" but how to maximize performance within it. The CPSWF family not only provides a precise language to frame this question but also offers the most suitable tools.

**Acknowledgment**

This work was supported by the Chinese Academy of Sciences (CAS) Project for Young Scientists in Basic Research (No. YSBR-049) and the Overseas Excellent Youth Science Foundation.

**Disclosures**

The authors declare no conflicts of interest.

**Data availability**

The codes and data used in this study will be available from the corresponding author upon reasonable request.

**Supplemental document**

See Supplementary Notes and Supplementary Videos 1-4 for supporting content.

**Supplementary Notes for**

Extreme Energy Concentration of Band-Limited Superoscillatory Vortices for Efficient Optical Micromanipulation

**Chengda Song,[1,2,3,#] Jing He,[1,2,3,4,#] Xi Xie,[5,*] Qian Wang,[4] Yijie Shen,[6,7,**] Fangwen Sun,[8,9] and Guanghui Yuan[1,2,3***]**

[1]Department of Optics and Optical Engineering, School of Physical Sciences, University of Science and Technology of China, Hefei, Anhui 230026, China

[2]State Key Laboratory of Opto-Electronic Information Acquisition and Protection Technology, Hefei, Anhui 230601, China

[3]Anhui Key Laboratory of Optoelectronic Science and Technologies, Hefei, Anhui 230026, China

[4]Institute of Materials Research and Engineering, Agency for Science, Technology and Research (A*STAR), 2 Fusionopolis Way, Innovis, Singapore 138634, Singapore

[5]School of Physics, Chengdu University of Technology, Chengdu 610059, China

[6]Centre for Disruptive Photonic Technologies, School of Physical and Mathematical Sciences, Nanyang Technological University, Singapore, Singapore

[7]School of Electrical and Electronic Engineering, Nanyang Technological University, Singapore, Singapore

[8]Chinese Academy of Sciences Center for Excellence in Quantum Information and Quantum Physics, University of Science and Technology of China, Hefei, Anhui 230026, China

[9]Hefei National Laboratory, University of Science and Technology of China, Hefei, Anhui 230088, China

[#]These authors contributed equally to this work

[*]xixie0721@163.com; [**]yijie.shen@ntu.edu.sg; [***]ghyuan@ustc.edu.cn

## Contents:

**SN 1. Physical origin and numerical computation of the CPSWFs**

CPSWFs are, at their physical core, the eigenmodes of a confocal laser cavity with a circular aperture of finite radius, where the aperture imposes a strict spatial band limit on the supported field. In Fig. S1(a), the relevant cavity geometry is defined such that the length $L$ and the cavity modes converge to those of the familiar Laguerre–Gaussian (LG) family, as shown in Fig. S1(b). But differently, LG modes can be thought of as the $a \to \infty$ limit of CPSWFs—they are what CPSWFs become when the band limit is removed. We stress that this does not imply that CPSWFs are somehow more fundamental than LG modes. The point is practical: whenever a physical band limit is present and cannot be ignored—whether it is set by a finite NA, an aperture stop, a maximum propagating wavevector, or a finite detection window—CPSWFs are the natural and appropriate basis for describing the field. Additionally, band-limited versions of other cavity mode families, such as band-limited Ince–Gaussian modes, remain largely unexplored and represent a worthwhile direction for future investigation.

There is an equally illuminating result in another asymptotic behaviour for $a \to 0$, where the LG, Bessel, and CPSWF modes all share the same l-order behavior near the origin:

$$E(r) \approx r^{|l|} \tag{S1}$$

The $l = 0$ member of this family is precisely the Airy disk produced by focusing a uniform flat-top beam. Extending the pattern to arbitrary $l$ defines a natural set of asymptotic modes, which we call Airy-Disk (AD) modes:

$$\mathrm{AD}_l(r) \propto \frac{J_{l+1}(r)}{r} \tag{S2}$$

Beyond these two limiting cases, a useful perspective on why band limits matter comes from geometrical optics. The conservation of étendue[1] — which holds at a level more fundamental than wave diffraction — already rules out the possibility of focusing an extended source to an infinitesimally small point. From a thermodynamic standpoint, the diffraction limit and energy efficiency are intrinsically linked: you cannot concentrate energy without paying a cost in étendue. What CPSWFs do is make this trade-off quantitative and optimal: given a finite circular target region (the FoV, which sets the band limit), they provide the field decomposition that maximizes the fraction of energy concentrated within that region.

Taken together, these perspectives reveal the band-limit parameter in CPSWFs not as an engineering choice but as a fundamental quantity intrinsic to any linear, band-limited signal system.

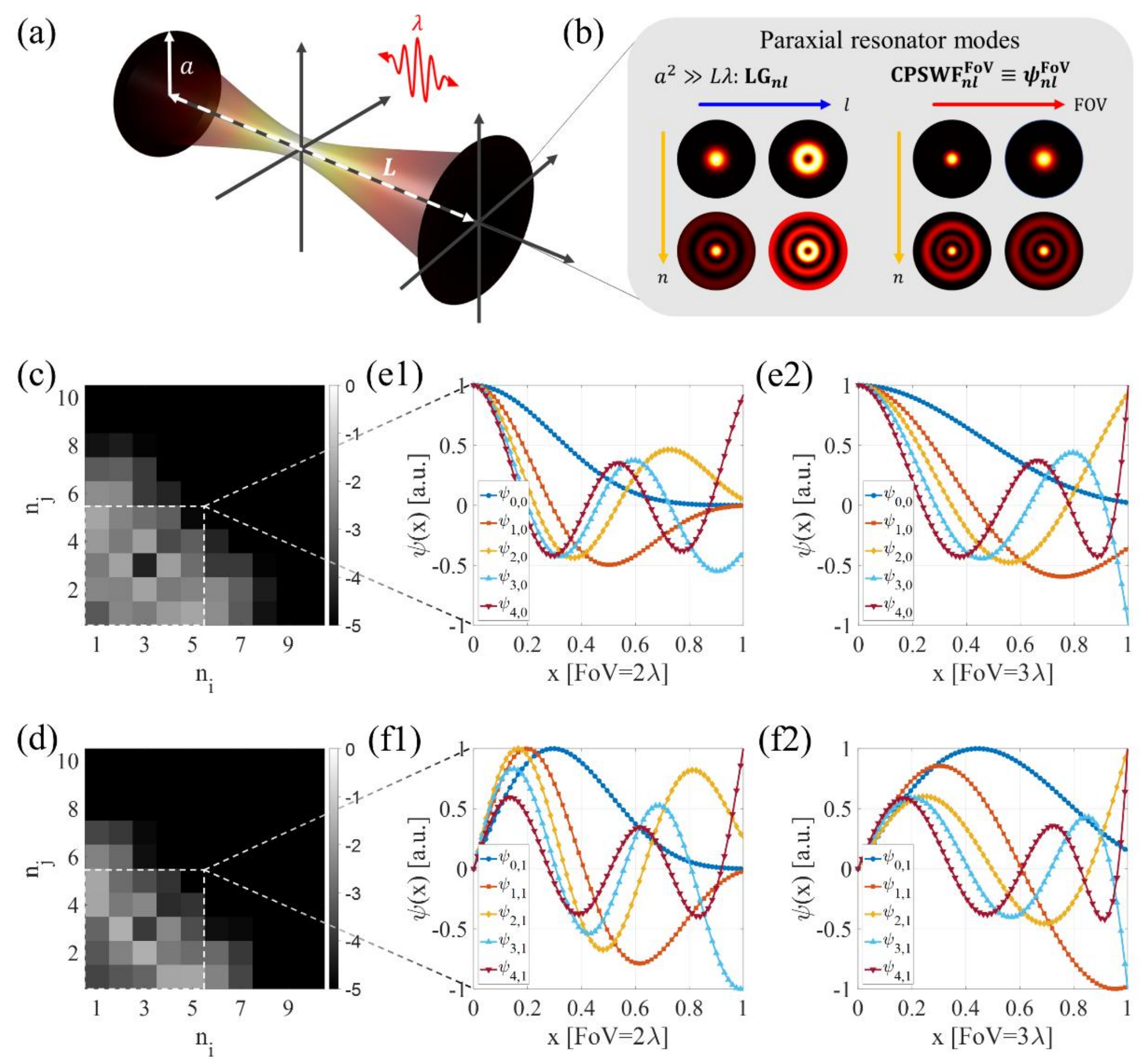

**Figure S1| Physical origin and numerical calculation of the CPSWF.** (a) A confocal laser cavity with finite circular aperture $a$ and cavity length $L$ (two times the focal length $R$), within which (b) the generated modes are shifted from LG modes to strictly band-limited CPSWF modes, with an emergent Spatial Band-limit Product (SBP) parameter $c = \mathrm{FoV} \cdot k$. The power spectra (unit: dB) of the finite Hankel transform with (c) $l = 0$ and (d) $l = 1$. The eigenproblem of the spectra matrix yields (e1) the first five n-ordered $\psi_{n,0}$ and (f1) the first five n-ordered $\psi_{n,1}$ at $\mathrm{FoV} = 2\lambda$. For completeness, (e2) and (f2) also show those of $\mathrm{FoV} = 3\lambda$.

Before CPSWFs can be put to use, they must of course be computed numerically. This is nontrivial, and it is fair to say that the difficulty of the computation has historically been one reason why CPSWFs are far less familiar than their one-dimensional counterparts, prolate spheroidal wave functions (PSWFs), which arise from analogous problems with a rectangular aperture. Our implementation follows the approach of Karoui and Moumni[2], who recognized that the band-limited Hankel transform operator can be discretized on a specially chosen orthogonal basis $T_{m,l}$, reducing the eigenmode problem to a standard matrix diagonalization. Figs. S1(c) and S1(d) show the resulting matrix form of the Hankel operators for $l = 0,1$, $\mathrm{FoV} = 2\lambda$; the first five eigenfunctions of each are shown in Figs. S1(e1) and S1(f1). For completeness, Figs. S1(e2) and S1(f2) show the same modes recomputed at different $\mathrm{FoV} = 3\lambda$. A key observation is that changing the FoV does not simply rescale the mode profiles—it fundamentally alters their structure, which underscores how the band limit is a genuinely nontrivial parameter.

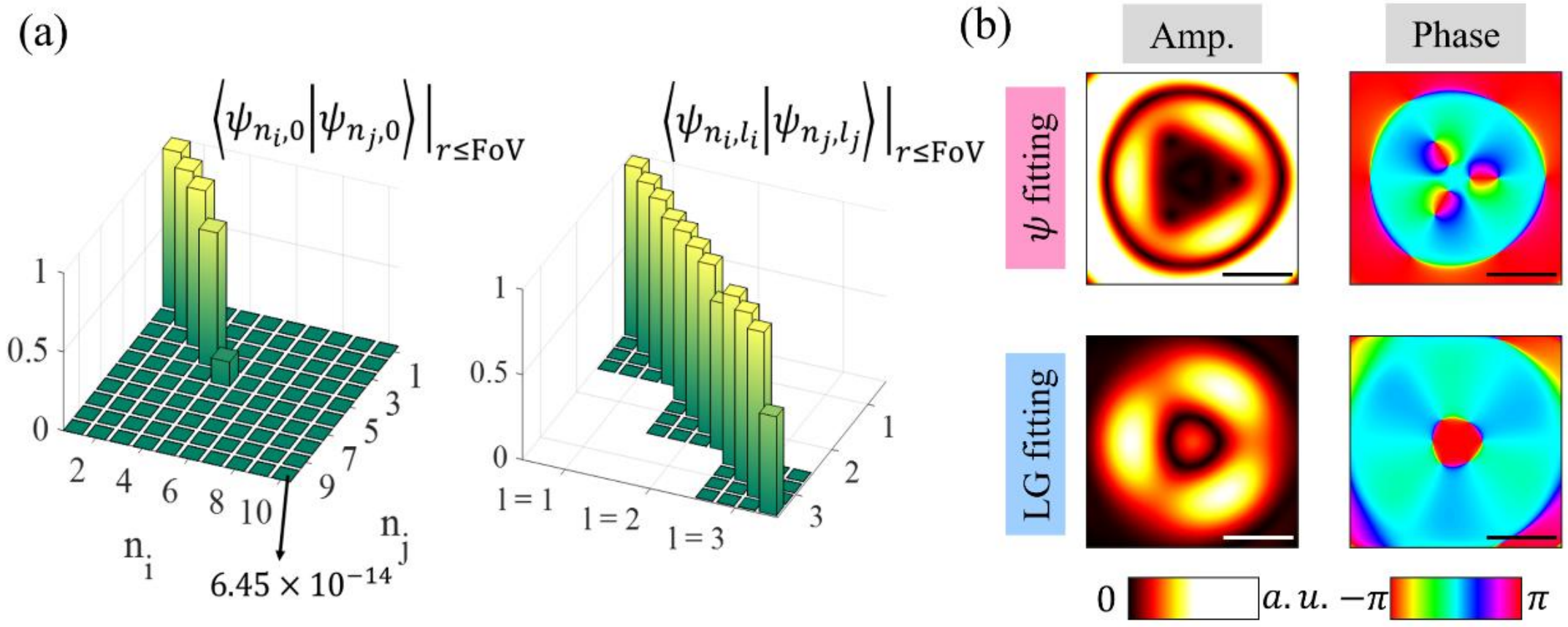

**Figure S2| CPSWFs as the basis sets for fitting complex optical fields.** (a) The orthogonality of $\psi_{n,0}$. The diagonal values represent the energy concentration within a predefined FoV ($1.5\lambda$), which decays exponentially from 1 to a vanishingly small energy (left panel). Consequently, only the first few n orders are physically practical for constructing the basis sets of CPSWFs with energy concentrations above 50% (right panel). (b) Amplitude and phase of the fitting results for both the CPSWF basis (top row) and the LG basis (bottom row). CPSWFs successfully retrieved the predefined trefoil knot field, with a deeply sub-diffraction, nontrivial topological structure of phase singularities. Scale bar: $1\lambda$.

As stated in the main text, CPSWFs are orthogonal on both [0, 1] and [0, +∞], but their eigenvalues decay exponentially with order $n$, indicating that only the first few modes carry an appreciable fraction of their energy within the designated FoV. This is illustrated in the left panel of Fig. S2(a), which shows the inner-product matrix for $\mathrm{FoV} = 1.5\lambda$, $l = 0$: the diagonal entries, each giving the in-FoV energy fraction for one mode, decay exponentially. Physically, there is a maximum number of nodes that can be packed into a region of fixed size; beyond this, the field necessarily spills outside the boundary, reducing energy efficiency.

In a practical sense, for any given band limit, only a finite subset of $(n, l)$ modes can serve as a useful basis for field synthesis—those whose in-FoV energy fraction exceeds a reasonable threshold, as illustrated in the right panel of Fig. S2(a). Fortunately, this restricted basis is nonetheless sufficient to reproduce highly complex optical fields. We demonstrate this with one explicit example.

We take as a target the trefoil optical knot field, a structured field whose phase singularities form a topologically nontrivial linked 3D texture. Following a seminal work from Dennis[3], the trefoil knot field at $z = 0$ is constructed by expanding the Milnor polynomial in the LG basis:

$$E_{tar}(r,\phi) = 1.51\ \mathrm{LG}_{0,0}(w) - 5.06\ \mathrm{LG}_{1,0} + 7.23\ \mathrm{LG}_{2,0} - 2.03\ \mathrm{LG}_{3,0} - 3.97\ \mathrm{LG}_{0,3}e^{i3\phi} \quad \text{(S3)}$$

For brevity, the beam waist $w$ is written explicitly only in the first term and is omitted from the remaining term. Rescaling $w$ simply enlarges or shrinks the entire field pattern, and the role of the band limit is precisely to impose a lower bound on how small $w$ can be.

For example, when $w = 0.5\lambda$ and NA=0.90, the k-space spectrum of the target field extends beyond the propagating-wave cutoff $\mathrm{NA} \cdot k_0$. LG modes, being unbounded, cannot faithfully represent such a field and act as a low-pass filter, and the topological structure of the phase singularities is lost. In contrast, CPSWFs are constructed precisely for this situation: the band limit is built into their definition, and they can represent the target field within the specified FoV. The reconstruction results are shown in Fig. S2(b). The CPSWF expansion faithfully recovers the nontrivial singularity topology, with singularity separations of approximately $0.39\lambda$ $(0.35\lambda/\mathrm{NA})$, whereas the LG expansion fails to reproduce the singularity structure.

This example concisely illustrates both the physical significance of band limits at subdiffraction scales and the unique capabilities of CPSWFs.

**SN 2. Extension to high-NA focusing**

**Vectorial focusing characteristics and deviations from paraxial approximations**

CPSWFs are derived within the scalar paraxial approximation, so a natural question is how well their properties are preserved under high-NA focusing, where polarization coupling becomes significant. This is the same test that other scalar mode families—LG, Hermite–Gaussian, and Bessel–Gaussian—have been subjected to, and we apply it here to CPSWFs.

We select two representative modes, one for each of the two key CPSWF attributes. For energy concentration, we choose $\psi_{0,1}^{\mathrm{FoV}=0.7\lambda/\mathrm{NA}}$, which is within the paraxial limit and is confined to more than 85% of the total energy within the subwavelength FoV. For superresolution, we choose $\psi_{4,0}^{\mathrm{FoV}=2.3\lambda/\mathrm{NA}}$, whose paraxial FWHM is approximately $0.362\lambda/\mathrm{NA}$. We examine whether these characteristics hold up under tight focusing at high NA. A side note on the first mode: $\psi_{0,1}^{\mathrm{FoV}=0.7\lambda/\mathrm{NA}}$ has a hollow transverse profile, which makes it compatible with cylindrical vector beams at the pupil—including azimuthal (AP) and radial (RP) polarizations, both of which require zero on-axis intensity.

The radial intensity profiles at the focal plane for $\psi_{0,1}^{\mathrm{FoV}=0.7\lambda/\mathrm{NA}}$ incident at NA = 0.9 are shown in Fig. S3(a1), with the 2D focal distributions shown in the inset. As expected for tight focusing, the results are polarization dependent. Both x-polarized and radially polarized (RP) incidence produce a bright central spot: in both cases, the spin–orbit interaction (SOI) in the +1-charge circular component generates a longitudinal field that has no vortex charge, and at high NA, this longitudinal component dominates, filling in the center.

Left-circular (LCP) and azimuthal (AP) incidence, by contrast, both preserve a hollow focal profile. With respect to the LCP, the SOI produces a longitudinal component with a +2 topological charge, keeping the center dark but at the cost of an expanded effective mode area. With respect to the AP incidence, the longitudinal component vanishes exactly; thus, no energy is diverted into a larger-area longitudinal field, which explains why the AP incidence achieves the best energy efficiency among the polarization states considered. It also suggests that AP-polarized CPSWFs can be promoted to a self-transforming vectorial eigenmode family, a point we would explore below.

The corresponding results for $\psi_{4,0}^{\mathrm{FoV}=2.3\lambda/\mathrm{NA}}$ at NA = 0.9 are shown in Fig. S3(a2). Under both linear and circular polarization incidence, the focal FWHM remains close to the scalar design value of $0.362\lambda/\mathrm{NA}$, with circular polarization maintaining circular symmetry in the focal spot.

We note that the transverse intensity is relevant for label-free microscopy. In fluorescence imaging, the random orientation of fluorophores means that all three polarization components contribute to excitation; thus, the total intensity must be considered instead. Under LCP incidence, the contribution of the longitudinal $E_z$ component decreases the total-intensity FWHM to approximately $0.42\lambda/\mathrm{NA}$. This is not a fundamental barrier: switching to a higher-order RP mode can recover a deep sub-diffraction total-intensity focus suitable for fluorescence-based superresolution imaging.

Figs. S3(b) and S3(c) systematically quantify how the two CPSWF advantages decrease as the NA increases. The energy concentration is acceptable. Even in the worst case—the LCP incidence—it decreases from the paraxial value of 86% to 77% at NA = 0.95, a relative decrease of less than 10%. The AP incidence remains above 85% across all NAs. The superresolution FWHM (evaluated for the LCP to maintain circular symmetry) is more robust: at NA = 0.95, it is approximately $0.365\lambda/\mathrm{NA}$, just 0.8% above the paraxial design value. For reference, this is essentially indistinguishable from the Bessel-beam diffraction limit — the difference being that Bessel beams carry infinite energy and thus cannot simultaneously define limits together with achievable energy.

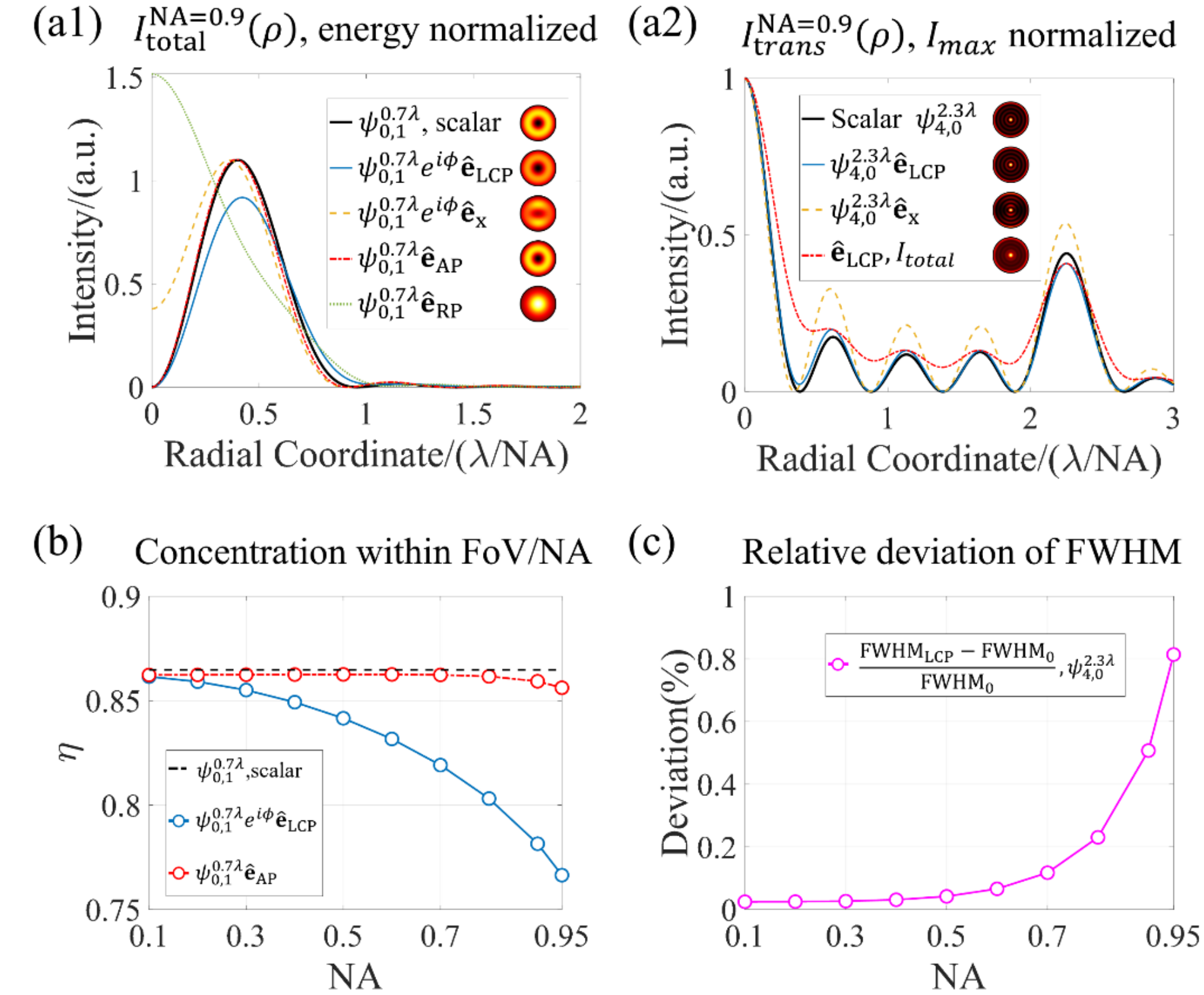

**Figure S3| Vectorial focusing characteristics and degeneration study against paraxial results**. Radial profiles of (a1) normalized total intensity for concentrator mode $\psi_{0,1}^{0.7\lambda}$ and (a2) normalized transverse intensity for the superresolving mode $\psi_{4,0}^{2.3\lambda}$ under NA = 0.9. Insets: corresponding 2D focal profiles. (b) Energy concentration $\eta$ within the FoV with increasing NA; black dashed line: scalar results, representing the upper boundary; blue: incident LCP vortex; red: incident AP. $\eta$ decreases with increasing NA (86.5% paraxial vs. 85.6% and 76.6% at NA=0.95), as expected. (c) Relative deviation of the FWHM under LCP illumination across the NA. Magenta curve: deviation from the scalar-designed FWHM. The maximum deviation reaches 0.81% at NA=0.95, validating the applicability of the paraxial design even under high NA.

In short, CPSWFs retain their defining advantages—efficient energy concentration and subdiffraction focal size—under high-NA tight focusing, with only modest, well-quantified polarization-dependent deviations from the paraxial predictions. This provides a solid basis for applying CPSWFs in high-NA settings such as superresolution microscopy, optical trapping, and nanofabrication, where both energy efficiency and focal quality matter.

**Vectorial self-transforms**

A separate but related question is whether one can find eigenmodes in a strict vectorial sense—fields that are self-reproducing under tight focusing not only in amplitude distribution but also in the polarization state.

High-NA focusing by an aplanatic lens is governed by the Richards–Wolf integral, which can be understood as a two-step process (Fig. S4(a)): First, the incident field is projected onto the refractive spherical wavefront via a polarization projection matrix $\mathbf{M}$; then, the resulting field is brought to focus by a Fourier transform. In Cartesian coordinates, the projection matrix takes the form:

$$\mathbf{M}(\theta,\phi)=\frac{1}{\sqrt{\cos\theta}}\begin{pmatrix}\cos^2\phi\cos\theta+\sin^2\phi & (\cos\theta-1)\sin\phi\cos\phi & -\sin\theta\cos\phi\\ (\cos\theta-1)\sin\phi\cos\phi & \cos^2\phi+\sin^2\phi\cos\theta & -\sin\theta\sin\phi\\ \sin\theta\cos\phi & \sin\theta\sin\phi & \cos\theta\end{pmatrix}\qquad\text{(S4)}$$

where $\theta=\operatorname{asin}\frac{\mathrm{NA}}{\mathrm{n}}\frac{x_p^2+y_p^2}{R_0}$ and $\phi=\operatorname{atan}\frac{y_p}{x_p}$ are spherical coordinates on the refractive surface, corresponding to the pupil coordinates $(x_p,y_p)$. For an azimuthally polarized (AP) incident field $\mathbf{E}_{\mathrm{inc}}^{\mathrm{AP}}=(-\sin\phi\,E_0(R),\cos\phi\,E_0(R),0)^{\mathrm{T}}$, a straightforward calculation shows that the focused field is also azimuthally polarized: $\mathbf{E}_{\mathrm{focal}}^{\mathrm{AP}}=\left(-\sin\varphi\,I_{1,-\frac{1}{2}}(r),\cos\phi\,I_{1,-\frac{1}{2}}(r),0\right)^{\mathrm{T}}$, where the radial integral is defined as follows:

$$I_{1,-\frac{1}{2}}(r)=\int_0^1\frac{E_0(k)}{\left(1-\left(\frac{\mathrm{NA}}{n}\right)^2\right)^{\frac{1}{4}}}J_1\left(k_0\frac{\mathrm{NA}}{\mathrm{n}}\cdot\mathrm{FoV}\cdot kr\right)kdk\qquad\text{(S5)}$$

Here, we define $k=\frac{R}{R_0}$ to connect directly with the main-text notation. In the paraxial limit, the factor under the square root approaches unity, and with an appropriate change in variables, Eq. (S5) reduces exactly to the scalar CPSWF equation given in the main text.

The radial intensity profiles before and after tight focusing for the AP eigenmode at NA = 0.9 (FoV $=0.7\lambda$) are shown in Fig. S4(b). The two profiles coincide, confirming a genuine vectorial self-transform.

An analogous self-transform exists for radially polarized (RP) incidence, provided that the longitudinal $E_z$ component is excluded. Together, the AP and RP eigenmode families form two orthogonal polarization channels, from which a rich variety of self-reproducing vector fields can be constructed. This confirms that the extension from scalar to vectorial CPSWFs is physically meaningful.

Beyond being a generalization of the scalar CPSWF equation, Eq. (S5) has its own significance: it directly encodes the NA as a tunable parameter, enabling mode designs that are tailored to a specific combination of focusing geometry and target field distribution—a degree of flexibility not available in the purely scalar theory.

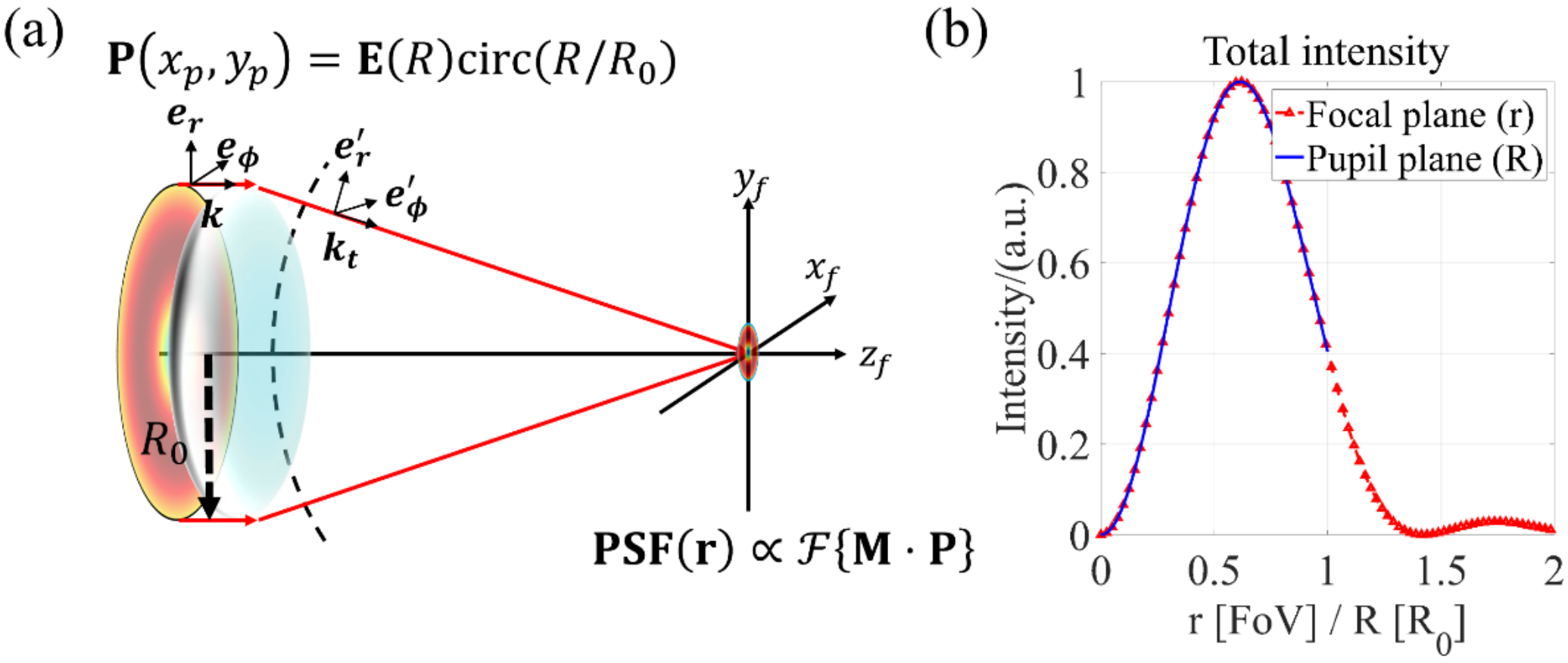

**Figure S4| Vectorial self-transforms**. (a) Schematic of high-NA focusing. The incident vectorial field $\mathbf{E}$ together with the finite aperture give the pupil function, which then undergoes two processes: (i) polarization projection onto the refractive spherical wavefront (by multiplying a projection matrix $\mathbf{M}$) and (ii) focusing toward the target plane (through Fourier transform). (b) A rigorous AP self-transform under a given pair of $\mathrm{NA} = 0.9$ and $\mathrm{FoV} = 0.7\lambda$.

## SN 3. Complete expression of optical forces of Rayleigh particles

A complete expression of optical forces on Rayleigh particles is as follows[4-6]:

$$\langle\mathbf{F}\rangle = \nabla\mathrm{U} + \frac{\sigma n}{c_0}\langle\mathbf{P}\rangle - \mathrm{Im}(\chi)\nabla\times\langle\mathbf{P}\rangle - \frac{\sigma_e c_0}{n}\nabla\times\langle\mathbf{S_e}\rangle - \frac{\sigma_m c_0}{n}\nabla\times\langle\mathbf{S_m}\rangle + \omega_0\gamma_e\langle\mathbf{S_e}\rangle + \omega_0\gamma_m\langle\mathbf{S_m}\rangle + \frac{c_0 n^3 k_0^4}{12\pi}\mathrm{Im}(\alpha_e\alpha_m^*)\mathrm{Im}(\mathbf{E}\times\mathbf{H}^*) \quad \text{(S6)}$$

where $\mathbf{E}$ and $\mathbf{H}$ are electromagnetic field vectors and $\mathrm{U} = \mathrm{Re}(\alpha_e)|\mathbf{E}|^2/4 + \mathrm{Re}(\alpha_m)|\mathbf{H}|^2/4 + \mathrm{Re}(\chi)\mathrm{Im}(\mathbf{E}\cdot\mathbf{H}^*)/2$ is the optical potential related to intensity. $\alpha_e$, $\alpha_m$, and $\chi$ are the electric, magnetic, and chiral polarizabilities of the particle, respectively. $\langle\mathbf{P}\rangle$ is the time-averaged Poynting vector, the lateral components of which are shown in Fig. S5(a), corresponding to the settings of Fig. 5 in the main text. $\langle\mathbf{S_e}\rangle$ and $\langle\mathbf{S_m}\rangle$ are the time-averaged electric and magnetic SAM densities, respectively. The interaction parameters $\sigma$, $\sigma_e$, $\sigma_m$, $\gamma_e$ and $\gamma_m$ represent light-beam cross sections, all of which are complex functions dependent on the particle properties (one can find their full expressions in Ref. [4]). The constants $n$, $c_0$, $\omega_0$, and $k_0$ denote the refractive index of the medium, the speed of light in vacuum, the frequency of light, and the wavenumber of light in vacuum, respectively.

The optical force on the particle (under the conditions of Fig. 5 in the main text) is decomposed into all eight terms of Eq. (S6) in Fig. S5(c). Three terms dominate among all of the forces: the gradient force (Fig. S5(c1)), the radiation pressure (Fig. S5(c2)), and the electric curl force (Fig.

S5(c4)); the remaining terms are negligible under these conditions. This justifies retaining only these three terms in the main-text expression.

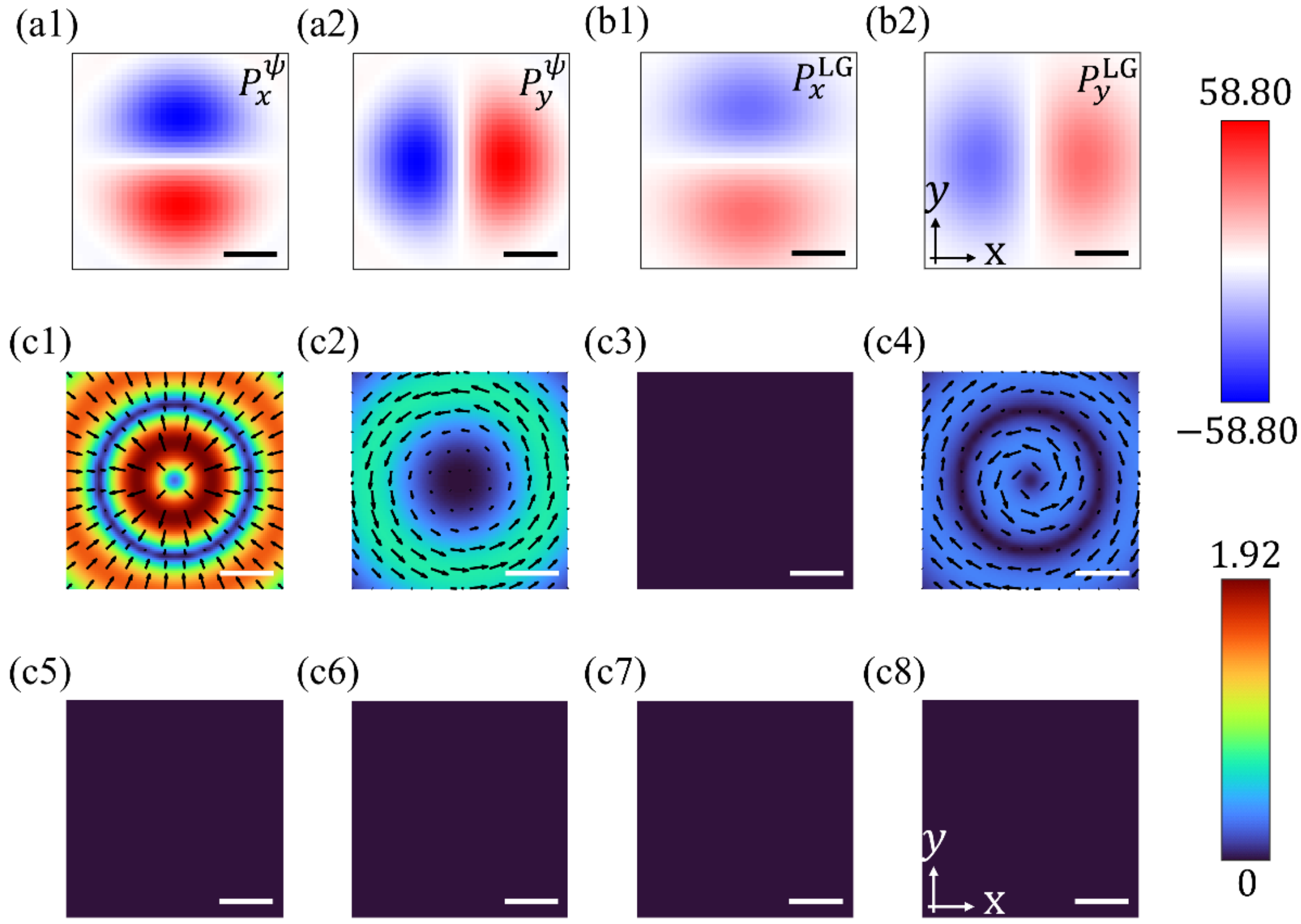

**Figure S5| Lateral Poynting vectors and optical forces**. (a1)–(a2) x- and y-components of the lateral Poynting vectors of the tightly focused $\psi_{0,1}$ mode. Scale bar: 0.2 μm. Unit: $\mathrm{mW/\mu m^2}$. (b1)–(b2) x- and y-components of the lateral Poynting vectors of the tightly focused $\mathrm{LG}_{0,1}$ mode. Scale bar: 0.2 μm. Unit: $\mathrm{mW/\mu m^2}$. (c1)–(c8) All eight terms of the optical induced forces of the mode $\psi_{0,1}$ according to Eq. (S6). Scale bar:0.2 μm. Unit: pN.

One practical consideration is that energy transport along the optical axis generates a nonzero axial force component. In a standard optical tweezer setup (Fig. S6(a)), this is typically handled by working near the coverslip surface, where the axial optical force is balanced by gravity and the supporting force from the coverslip[6,7]. Under these conditions, the lateral dynamics are effectively decoupled from the axial dynamics and can be studied independently without loss of generality. The longitudinal Poynting vector distributions for the CPSWF and LG modes are shown in Figs. S6(b1) and S6(b2), respectively, for the parameters of Fig. 4 in the main text.

A noteworthy feature of superoscillatory fields is that energy backflow[8]—the local propagation of energy in the direction opposite to the beam—can occur at certain off-axis positions. Such backflow may give rise to a transverse angular momentum oriented perpendicular to the optical axis.

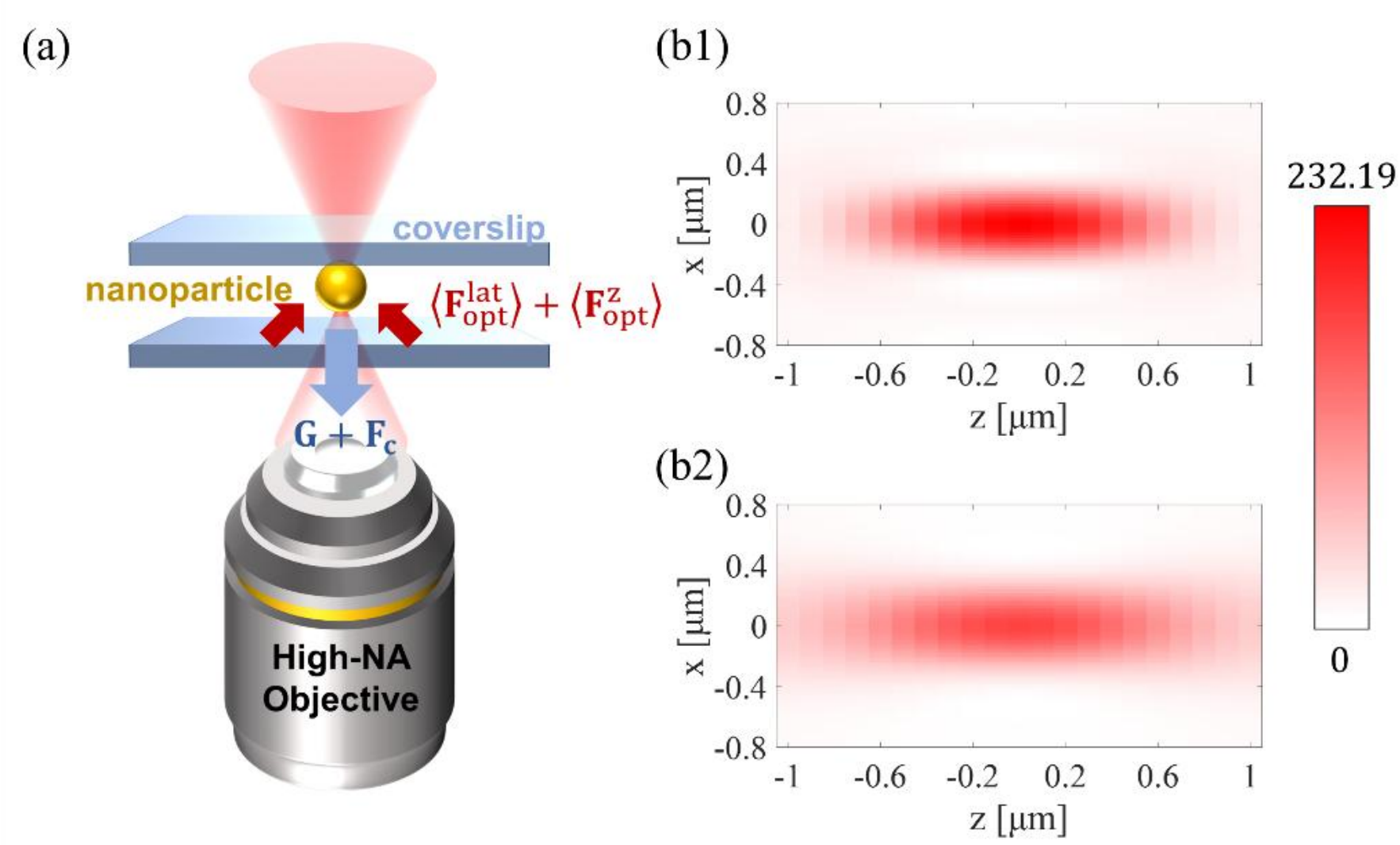

**Figure S6| Longitudinal forces.** (a) Schematic of a typical optical tweezer setup. (b1) Longitudinal component of Poynting vectors of the tightly focused $\psi_{0,1}$ mode. (b2) Longitudinal component of Poynting vectors of the tightly focused $\mathrm{LG}_{0,1}$ mode.

We now examine the robustness of the results in the main text to variations in the key experimental parameters.

Figs. S7(a1)–(a2) show how the radial and azimuthal force components vary with respect to the particle radius, with all the other parameters fixed at the main-text values (Au, NA = 1.2; main-text case marked with magenta pentagrams). Increasing the radius progressively enhances the azimuthal force relative to the radial force. This is because a larger scattering cross section amplifies the radiation pressure term in Eq. (S6), which is the dominant driver of the azimuthal force. From the explicit expressions for $\alpha_e$, $\alpha_m$ and $\sigma$, the peak radial and azimuthal forces scale as power laws with exponents 3 and 6, respectively: $F_{\max,rad} \sim 1.457 \cdot 10^{-5}$ [pN · nm$^{-3}$] $a^3$, $F_{\max,azi} \sim 3.838 \cdot 10^{-11}$ [pN · nm$^{-3}$] $a^6$, both with $R^2 > 0.95$. One might be tempted to use larger particles to obtain a stronger and purer azimuthal force, but this comes at the cost of invalidating the Rayleigh approximation ($a \ll \lambda$). The choice of 50 nm in the main text is a deliberate conservative compromise.

Figs. S7(b1)–(b2) compare the forces on a gold (Au) nanoparticle and a polystyrene (dielectric) particle of the same size. The dielectric particle experiences much weaker forces overall and almost no azimuthal force. The reason is straightforward: a dielectric particle is nearly non-absorbing, so the optical force is dominated by the momentum change of the scattered light rays, with negligible angular momentum transfer. A metal particle near its plasmon resonance, by contrast, absorbs

strongly, enabling efficient transfer of orbital angular momentum from the beam to the particle. Quantitatively, the azimuthal force in the Rayleigh regime is proportional to $\mathrm{Im}[\alpha]$, and the total scattering cross section differs by approximately three orders of magnitude between the two materials ($\sigma_{Au} = 4.67 \cdot 10^{-15}, \sigma_{(C_8H_8)_n} = 1.67 \cdot 10^{-18}$), which is consistent with the observed difference in the peak azimuthal force in panel (b2). We use Au at 800 nm specifically because the excitation wavelength lies near the plasmon resonance, maximizing the azimuthal force. The refractive indices of both materials at 800 nm follow those in Ref. [9] and [10].

Finally, Figs. S7(c1)–(c2) show the NA dependence. This trend is unsurprising: tighter focusing concentrates more energy into a smaller area, increasing the peak intensity gradient and hence the optical force. This is in fact analogous to the comparison between LG and CPSWF modes: at fixed NA, structured-light engineering with CPSWFs localizes the energy into a prescribed region more efficiently than LG modes can. In other words, CPSWFs effectively relaxes the NA requirement, which is a meaningful advantage in experiments where very high-NA objectives are impractical or unavailable.

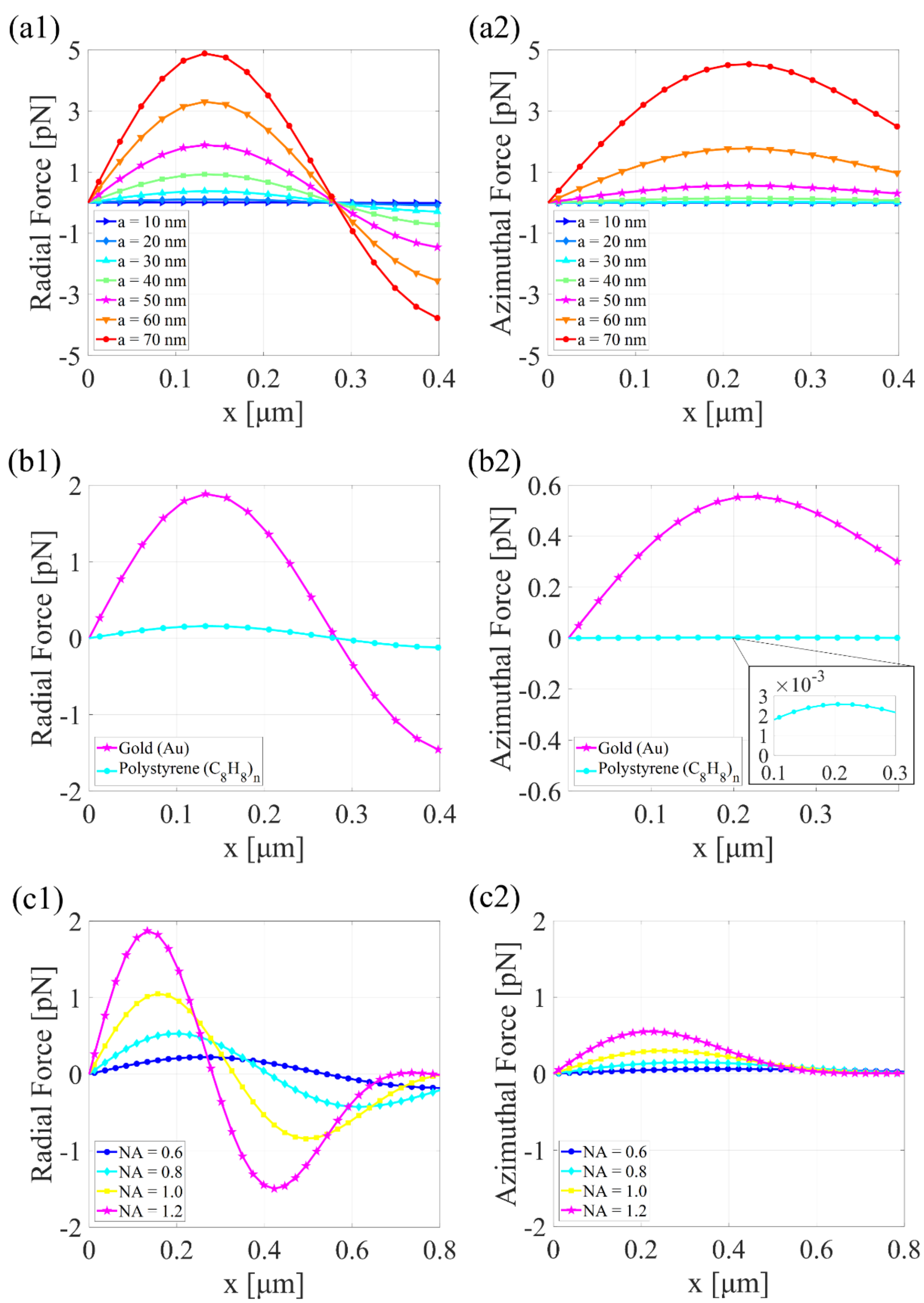

**Figure S7| Parameter generality.** (a1)–(a2) Radial and azimuthal forces exerted on Au nanoparticles with various radii; NA=1.2. (b1)–(b2) Radial and azimuthal forces exerted on metallic (Au) and dielectric (Polystyrene) nanoparticles with a radius of 50 nm and a NA=1.2. (c1)-(c2) Radial and azimuthal forces exerted on Au nanoparticles with a radius of 50 nm under different focusing NAs. Magenta lines with pentagram markers correspond to the settings of Fig. 5 in the main text.

**SN 4. Langevin dynamics of nanoparticles in a viscous flow**

The translational motion of a spherical nanoparticle in a viscous fluid is governed by the Langevin equation[6], which accounts for hydrodynamic drag, optical-induced forces, and stochastic thermal fluctuation:

$$\frac{d}{dt}\left(m_p \frac{d\mathbf{q}}{dt}\right) = \mathbf{F}_{\mathrm{D}} + \mathbf{F}_{\mathrm{opt}}(\mathbf{q}) + \mathbf{F}_{\mathrm{B}} \tag{S7}$$

where $\mathbf{q}$ denotes the particle position in the lateral xy-plane and $m_p$ is the particle mass with radius $a_p$. The first term, drag force $\mathbf{F}_{\mathrm{D}}$, follows the Stokes–Schiller–Naumann formulation:

$$\mathbf{F}_{\mathrm{D}} = \frac{1}{\tau_p} m_p \left(\mathbf{u}_0 - \frac{d\mathbf{q}}{dt}\right), \tau_p = \frac{4\rho_p \left(2a_p\right)^2}{3\eta C_D \mathrm{Re}_{\mathrm{r}}} \tag{S8}$$

where $\mathbf{u}_0$ is the preset background fluid velocity, $\tau_p$ is the relaxation time, and $\eta = 10^{-3}\ \mathrm{Pa \cdot s}$ is the dynamic viscosity chosen as that of water. $\mathrm{Re}_{\mathrm{r}} = \frac{\rho\left|\mathbf{u}_0 - \frac{d\mathbf{q}}{dt}\right| d_p}{\eta}$ is relative Reynolds number and $C_D = \frac{24}{\mathrm{Re}_r}\left(1 + 0.15\mathrm{Re}_r^{0.687}\right)$ is drag coefficient. For a radius under $100\ \mathrm{nm}$ and $|\mathbf{u}_0|$ under $1000\ \mu\mathrm{m/s}$, the Reynolds number is approximately $10^{-4}$ according to Eq. (S8), the Schiller–Naumann correction is well approximated by the standard Stokes drag, where $\mathbf{F}_{\mathrm{D}} = \gamma\left(\mathbf{u}_0 - \frac{d\mathbf{q}}{dt}\right)$ with $\gamma = 6\pi\eta a$.

The third term is the Brownian force, modelling stochastic thermal fluctuations:

$$\mathbf{F}_{\mathrm{B}} = \boldsymbol{\zeta}\sqrt{\frac{12\pi \mathrm{k_B} \eta T a}{\Delta t}} \tag{S9}$$

where $\boldsymbol{\zeta} = (\zeta_x, \zeta_y)$ is a vector of independent normally distributed random variables. $\mathrm{k_B}$ is the Boltzmann constant, $T = 293.15$ K is room temperature, and $\Delta t$ is the time step for solving the equations. This formulation is consistent with the fluctuation–dissipation theorem, with the recovery of the Stokes–Einstein diffusion coefficient $D = k_B T/\gamma$.

The numerical methods employed in COMSOL Multiphysics included a combination of fluid dynamics and particle tracking to monitor the particle trajectories within the defined domain of $[-4\lambda/n, 4\lambda/n]^2$. The simulation timestep was set to $\Delta t = 1$ ns, aligned with the relaxation time $\tau_p \sim 7\ \mathrm{ns}$ to ensure accurate capture of stochastic particle dynamics. To enhance spatial resolution in regions of high force gradient, adaptive meshing was employed with a minimum element size of $4 \times 10^{-5}\ \mu\mathrm{m}$, with the optical force field evaluated at each particle position through bilinear interpolation of the precomputed force map. Boundary conditions are set as diffuse scattering condition — upon collision, the particle speed is preserved and the reflected velocity is decomposed into a normal component directed inward and a tangential component along the boundary. Each simulation involves 41 particles, which are tracked simultaneously over a total duration of 5 ms.

**SN 5. Laser-induced heating**

When a metallic nanoparticle is illuminated by a focused laser beam, a fraction of the incident optical power is absorbed and converted to heat via Ohmic losses, generating a volumetric heat source $Q = \frac{1}{2}\mathrm{Re}\{\mathbf{J} \cdot \mathbf{E}^*\}$, which drives a local temperature rise governed by the heat equation[11]:

$$\rho c_p \frac{\partial T}{\partial t} = \nabla \cdot (\kappa \nabla T) + Q(\mathbf{r}, t) \tag{S10}$$

where $\rho$, $c_p$ and $\kappa$ are the mass density, specific heat capacity, and thermal conductivity of the medium, respectively. This coupled electromagnetic–thermal problem was solved in COMSOL Multiphysics, where the precomputed tightly focused CPSWF field was imported as complex Cartesian components $E_x$, $E_y$ and $E_z$, and used as the background optical field for a frequency-domain electromagnetic simulation. A spherical Au nanoparticle with a radius of 50 nm was placed in water at the representative trapping-ring position.

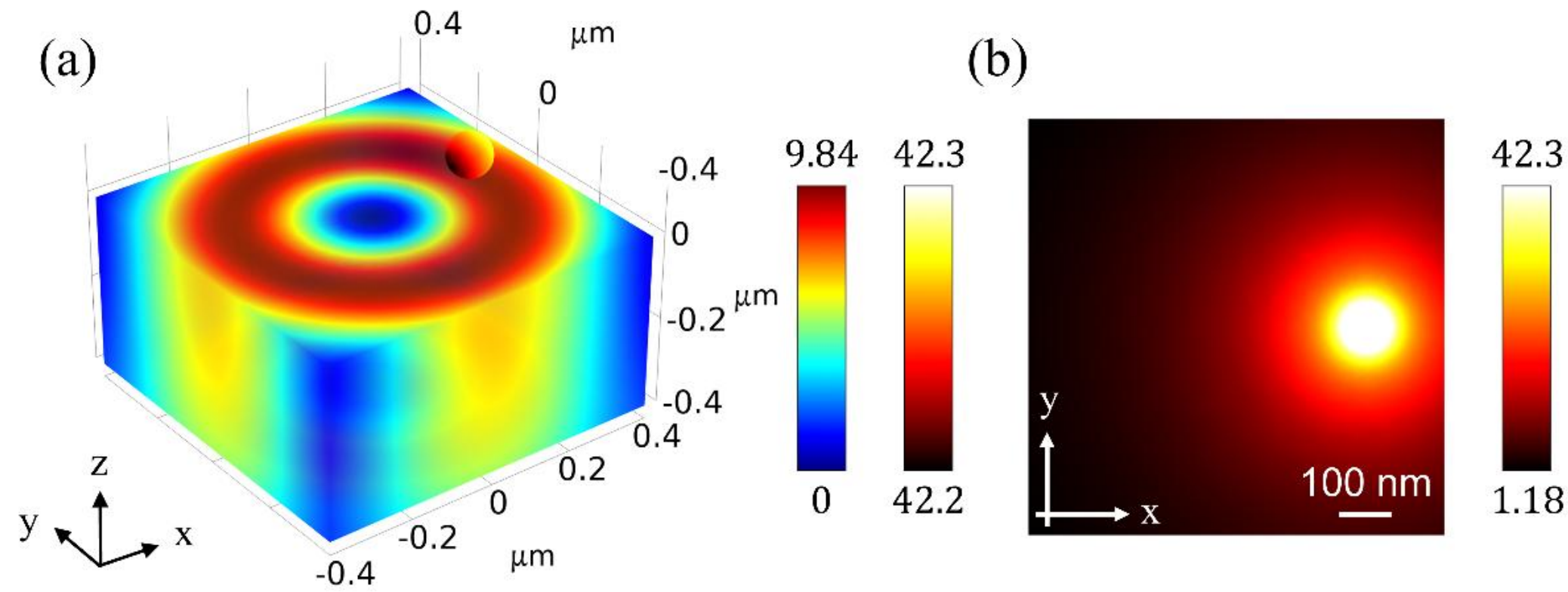

**Figure S8| Thermo effects of Au nanoparticle at resonance wavelength 800 nm.** (a) 3D electrical field distribution and local temperature rise of a trapped Au nanoparticle of a radius of 50 nm. The top surface reveals the annular focal-field intensity profile as plotted in main text Fig. 5. The left colorbar indicates the 3D intensity (unit: $\mathrm{V}^2 \cdot \mathrm{\mu m}^{-2}$) while the right colorbar corresponds to the local temperature rise, which is about 42 K. (b) Temperature rise in x-y plane, which remains moderate and would not disturb Langevin dynamics as discussed in main text.

The electromagnetic loss in the particle was then coupled to the Heat Transfer in Solids and Fluids module through electromagnetic heating. The optical wavelength was set to 800 nm, with water treated as the surrounding medium and Au assigned its complex refractive index at this wavelength.

The thermal simulation used temperature-dependent water properties, while the outer boundaries of the water domain were fixed at the ambient temperature of 293.15 K. These settings follow the uploaded COMSOL model, which imports the 3D focal-field data, solves the electromagnetic and heat-transfer modules, and evaluates the temperature rise as T−293.15 K.

As shown in Fig. S8(a), the trapped particle locates at the intensity maxima, thus reaching a steady-state temperature rise of ~42 K above ambient under the illumination conditions discussed in Fig. 5 (wavelength: 800 nm; power: 25 mW; mode: tightly focused $\psi_{0,1}^{0.7\lambda}e^{i\phi}$ under NA=1.2). This level of heating remains moderate and is not expected to induce significant thermophoretic drift, bubble nucleation or phase transition, confirming that photothermal effects do not fundamentally limit stable manipulation of Au nanoparticles at this power level.

**SN 6. Fairer and broader comparisons across varying beam modes.**

One might wonder if the comparison between LG and CPSWF in the main text shows rather narrow coverage, thus making the assertion "CPSWFs outperform any band-limited alternative" skeptical. As a result, we consider it urgent to compare it against other band-limited alternatives. For example, we choose three representatives—radially polarized higher order-LG[12], binary flat-top[13] (representing Toraldo-type annular filters) and numerically optimized pupil masks[14]—all of which have demonstrated solid improvements in either optical tweezer configurations or superresolution microscopy.

To compare CPSWFs against a broader set of aforementioned structured-light alternatives, we define four performance metrics guided by their most common figures of merit in optical trapping and superresolution applications:

**(a)** **FWHM** of the central focal spot. This is the standard figure of merit for resolving ability in point spread function engineering: a smaller FWHM indicates a higher spatial resolution. In trapping, it reflects the lateral confinement of the trapping potential.

**(b)** The peak intensity at the focal center, $\boldsymbol{I_c}$, is normalized to the power density (mW/μm²) at a fixed total incident power of 100 mW. A higher central intensity indicates a deeper trapping well and a better signal-to-noise ratio in superresolution imaging.

**(c)** The peak sidelobe-to-center intensity ratio $\boldsymbol{I_s/I_c}$. Lower is better. In single-lens imaging systems, strong sidelobes can severely degrade image quality; in confocal systems, they represent energy wasted on components that do not contribute to the image, even if a pinhole rejects them spatially.

**(d)** The trapping stiffness $\boldsymbol{\kappa}$, which is proportional to the restoring force near the trapping center. It captures both the depth and the confinement of the trapping well and can be regarded as a combined measure of metrics (**a**) and (**b**).

Table S1 lists these metrics for three representative published modes (rows 1–3) and four CPSWF modes matched to each (rows 4–7). The polarization state, NA, incident power, wavelength, and working distance are all matched between each pair. The CPSWFs consistently outperform the alternatives on every metric. The inset figures show the total intensity distributions in the lateral $[-4\lambda, 4\lambda]^2$ and axial $[-4\ \mu\text{m}, 4\ \mu\text{m}]^2$ planes. One caveat is worth noting for row 3 (azimuthally polarized binary spiral zone plate): the original paper[13] states that this element satisfies the tangential focus condition rather than the sinusoidal condition obeyed by the other systems. This means that the trapping stiffness comparison (metric **d**) is not directly applicable.

| Mode | Polarization | NA | $I_{xy}$ | $I_{xz}$ | FWHM | $I_c$ | $I_s/I_c$ | $\kappa$ |
|---|---|---|---|---|---|---|---|---|
| $LG_{3,1}$[11] | RP | 1.4 | | | 0.358 | 76.15 | 0.73 | 0.615 |
| Binary Flattop[12] | LCP | 0.55 | | | 0.255 | 5.34 | 4.88 | 0.0109 |
| Bessel× BSZP[13] | AP | 0.95 | | | 0.409 | 3.06 | 0.30 | - |
| $\psi_{3,1}^{1.87\lambda}$ | RP | 1.4 | | | 0.356 | 81.30 | 0.71 | 0.666 |
| $\psi_{2,1}^{1.2\lambda}$ | RP | 1.4 | | | 0.359 | 138.87 | 0.58 | 1.12 |
| $\psi_{1,0}^{0.36\lambda}$ | LCP | 0.55 | | | 0.255 | 7.11 | 4.69 | 0.0164 |
| $\psi_{1,1}^{1.5\lambda}e^{i\phi}$ | AP | 0.95 | | | 0.410 | 145.75 | 0.69 | 0.356 |

**Table S1| Broader comparison across varying beam modes[11-13], considering band limits.** Inset figures: total intensity distribution on both the lateral (xy-) and axial (xz-) planes.